\documentclass[english,aps,nofootinbib,preprint]{revtex4}
\usepackage[T1]{fontenc}
\usepackage[latin9]{inputenc}
\usepackage[active]{srcltx}
\usepackage{color}
\usepackage{amsmath}
\usepackage{amssymb}
\usepackage{graphicx}
\usepackage{esint}

\makeatletter
\@ifundefined{textcolor}{}
{%
 \definecolor{BLACK}{gray}{0}
 \definecolor{WHITE}{gray}{1}
 \definecolor{RED}{rgb}{1,0,0}
 \definecolor{GREEN}{rgb}{0,1,0}
 \definecolor{BLUE}{rgb}{0,0,1}
 \definecolor{CYAN}{cmyk}{1,0,0,0}
 \definecolor{MAGENTA}{cmyk}{0,1,0,0}
 \definecolor{YELLOW}{cmyk}{0,0,1,0}
}

\usepackage{babel}

\makeatother

\usepackage{babel}
\begin{document}

\title{Neutralino Dark Matter in Gauge Mediation After Run I of LHC and
LUX}

\author{Ran Ding $^{a}$}

\author{Liucheng Wang $^{b}$}

\email{lcwang@udel.edu}

\thanks{(Corresponding Author)}

\author{Bin Zhu $^{a}$}

\affiliation{$^{a}$School of Physics, Nankai University, Tianjin 300071, P. R.
China\\
$^{b}$Bartol Research Institute, Department of Physics and Astronomy,
University of Delaware, Newark, DE 19716, USA}
\begin{abstract}
Neutralino can be the dark matter candidate in the gauge-mediated
supersymmetry breaking models if the conformal sequestered mechanism
is assumed in the hidden sector. In this paper, we study this mechanism
by using the current experimental results after the run I of LHC and
LUX. By adding new Yukawa couplings between the messenger fields and
Higgs fields, we find that this mechanism can predict a neutralino dark
matter with correct relic density and a Higgs boson
with mass around 125 GeV. All our survived points have some common
features. Firstly, the Higgs sector falls into the decoupling limit.
So the properties of the light Higgs boson are similar to the predictions
of the Standard Model one. Secondly, the correct EWSB hints a relatively
small $\mu$-term, which makes the lightest neutralino lighter than
the lightest stau. So a bino-higgsino dark matter with correct relic
density can be achieved. And the relatively small $\mu$-term results
in a small fine-tuning. Finally, this bino-higgsino dark matter can
pass all current bounds, including both spin-independent and spin-dependent
direct searches. The spin-independent cross section of our points
can be examined by further experiments.
\end{abstract}
\maketitle

\section{Introduction}

It is now believed that the dominant matter in the universe should
be non-baryonic dark matter (DM) instead of visible ones. And DM should
not be composed of any known Standard Model (SM) particles. Extra
symmetry is usually necessary to make DM stable on the cosmological
time scale. In supersymmetric (SUSY) models, if the R-parity conservation
is assumed, the lightest supersymmetric particle (LSP) is absolutely
stable. The LSP should be a good DM candidate if it is electrically
neutral. On the other hand, the measurement of relic density generally
suggests that the DM mass is around several GeV to 10 TeV with a weak
interaction. That is to say, the LSP is expected to be a weakly interacting
massive particle (WIMP).

Unfortunately, gravitino with mass less than 1 GeV is usually the
LSP in the gauge mediation supersymmetry breaking (GMSB) models. GMSB
\cite{Dine:1981gu,Dine:1981za,Dimopoulos:1981au,Nappi:1982hm,AlvarezGaume:1981wy,Dine:1993yw,Dine:1993qm,Dine:1994vc,Dine:1995ag}
is one of the promising mechanisms to describe the SUSY-breaking in
the minimal supersymmetric Standard Model (MSSM) (for a modern review,
see \cite{Giudice:1998bp}). The effect of SUSY breaking is mainly
transmitted to the MSSM sector through the gauge interaction, which
makes GMSB models flavor safe. The soft masses from gravity mediation
are suppressed by Planck-scale and not generation-blind. 
So these Planck-scale induced soft masses are dangerous as they mediate flavor-changing effects. 
In order to escape from experimental constraints, these dangerous Planck-scale induced soft masses should be tiny. 
As the gravitino mass also arises from the Planck-scale induced operator, gravitino
is always the LSP in GMSB models. Such a gravitino DM is hard to be
detected and its relic density depends on the dynamics of inflation.
Generally speaking, the lack of the predictability of gravitino DM
is one of the drawbacks of GMSB models.

Instead of gravitino, the lightest neutralino can be the DM candidate
in GMSB models if the hidden sector is strongly coupled \cite{Craig:2008vs,Shirai:2008qt,Endo:2009uj,Craig:2009tz}.
The conformal sequestered hidden sector can raise the gravitino mass
relative to the dangerous Planck-scale induced soft masses without
introducing any flavor physics problems. As studied in \cite{Craig:2008vs,Shirai:2008qt,Endo:2009uj,Craig:2009tz},
neutralino DM in the gauge mediation with sequestered SUSY breaking
is typically purely bino-like and its mass is within the WIMP range.
Since neutralino is the LSP, the lightest tau slepton (stau) should
be heavier than the lightest neutralino. This is a strong constraint
to those models, which requires the messenger scale $M_{\mathrm{mess}}$
typically around $10^{10}$ GeV. Compared to low-scale gauge mediation,
stau will be heavier in such a high-scale gauge mediation, as the
stau mass grows up when renormalization group equations (RGEs) of
MSSM are running from the input scale down to the electroweak scale.

All above papers about neutralino DM in GMSB scenarios with sequestered
SUSY breaking were done several years ago. After the run I of Large
Hadron Collider (LHC) and Large Underground Xenon (LUX) DM experiment,
these models are necessary to be revisited and carefully checked by
current experimental constraints. Firstly, a SM-like Higgs boson with
mass around 125 GeV has been confirmed at LHC \cite{Aad:2012tfa,Chatrchyan:2012ufa}.
A 125 GeV Higgs in decoupling MSSM scenario prefers either a heavy
top squark (stop) or a large $A_{t}$-term \cite{Carena:2011aa,Hall:2011aa,Draper:2011aa,Baer:2011ab,Li:2011ab,Heinemeyer:2011aa,Arbey:2011ab,Kang:2012sy,Cao:2012fz,Ajaib:2012vc,Ke:2012qc},
since both could contribute large loop corrections to the Higgs mass.
Unfortunately, minimal GMSB models predict vanishing $A$-terms at
the messenger scale, which presents another challenge for GMSB models.
Secondly, no signals of SUSY particles have been detected at LHC. 
Together with a 125 GeV Higgs, it raises uncomfortable issues with
naturalness which are widely discussed in \cite{Barbieri:1987fn,Anderson:1994dz,Cohen:1996vb,Ciafaloni:1996zh,Bhattacharyya:1996dw,Chankowski:1997zh,Barbieri:1998uv,Kane:1998im,Giusti:1998gz,BasteroGil:1999gu,Feng:1999mn,Romanino:1999ut,Feng:1999zg,Chacko:2005ra,Choi:2005hd,Nomura:2005qg,Kitano:2005wc,Nomura:2005rj,Lebedev:2005ge,Kitano:2006gv,Allanach:2006jc,Giudice:2006sn,Perelstein:2007nx,Allanach:2007qk,Cabrera:2008tj,Cassel:2009ps,Barbieri:2009ev,Horton:2009ed,Kobayashi:2009rn,Lodone:2010kt,Asano:2010ut,Strumia:2011dv,Cassel:2011tg,Sakurai:2011pt,Papucci:2011wy,Larsen:2012rq,Baer:2012uy,Espinosa:2012in,Boehm:2013qva,Zheng:2013afa,Zheng:2013kwa}.
Finally, the updated bounds of DM direct searches become severer than
the bounds in previous studies. The current strictest bound is given
by the LUX Collaboration \cite{Akerib:2013tjd}, who is the first
to break the $10^{-45}\;\mathrm{cm^{2}}$ cross section barrier of
DM spin-independent detection at some WIMP mass range. New LUX upper
limits have already been used to constrain DM in SUSY models \cite{Cao:2013mqa,Ellis:2013oxa,Buchmueller:2013rsa,Martin:2013aha,Guo:2013asa}.
All in all, in this paper we would focus on these new constraints
on GMSB models with sequestered SUSY breaking.

This paper is organized as follows. In Section II, we give a brief
review about the GMSB scenarios with sequestered SUSY breaking
and how to get a neutralino DM in GMSB models. Section III is devoted
to studying new constraints on those GMSB models and showing our results.
We finally conclude with a summary in Section IV.

\section{Gauge mediation with sequestered SUSY breaking}

In this section, we give a brief review about the GMSB models with
the sequestered SUSY breaking and how to get a neutralino DM. We start
with the minimal GMSB model. As a singlet superfield $S$ in the hidden
sector breaks SUSY, the messenger superfields $\Phi$ couple to the
hidden field $S$ via a superpotential $W=\kappa S\Phi\bar{\Phi}$
with $\kappa\sim\mathcal{O}(1)$%
\footnote{Because of $\mathcal{O}(1)$, $\kappa$ is neglected in many papers
for simplify. %
}. In the view of a spurion field, $S=\left\langle s\right\rangle +F_{s}\theta^{2}$
is assumed to parameterize the typical effect of SUSY breaking. As
a low-energy effective field theory of SUSY, many higher-dimensional
operators contribute to the K\"ahler potential after heavy fields are
integrated out. Sfermions get soft masses through the following operators
\begin{equation}
\mathcal{K}_{\mathrm{eff}}=\frac{S^{\dagger}S}{M_{\mathrm{mess}}^{2}}\underset{i}{\sum}c_{i}F_{i}^{\dagger}F_{i}+\frac{S^{\dagger}S}{M_{PL}^{2}}\underset{i,j}{\sum}b_{i,j}F_{i}^{\dagger}F_{j},\label{eq:kalher}
\end{equation}
where $F_{i}$ are superfields of sfermions in the visible sector.
The messenger scale is $M_{\mathrm{mess}}=\kappa\left\langle s\right\rangle $
and $M_{PL}$ is the Planck scale. Since $M_{\mathrm{mess}}\ll M_{PL}$
in GMSB models, the soft masses $m^{\mathrm{soft}}$ mainly come from
the first term of Eq.(\ref{eq:kalher}), which are $ $proportional
to $\frac{NF_{S}}{M_{\mathrm{mess}}}$. Here $N$ is the effective
number of the messenger fields. Because the gauge interaction is flavor-blind,
$M_{\mathrm{mess}}$-scale induced operators naturally escape from
experimental constraints on the flavor violation. However, the Planck-scale
induced operators are very dangerous since the Wilson coefficients
$b_{i,j}$ are not diagonal under the flavor index $i$, $j$ of the
sfermions. Since $b_{i,j}$ are always expected to be $\mathcal{O}(1)$,
the Planck-scale induced soft masses are $m_{PL}^{\mathrm{soft}}\sim\frac{F_{S}}{M_{PL}}\sim m_{3/2}$.
In order to avoid the flavor problems at electoweak scale, $m_{PL}^{\mathrm{soft}}$
have to be less than 1 GeV. That is why gravitino is always the LSP
in GMSB models.

However, the dynamics of the hidden sector may be important to determine
the MSSM spectrum if the SUSY breaking sector is strongly coupled
\cite{Luty:2001jh,Luty:2001zv,Dine:2004dv,Cohen:2006qc,Schmaltz:2006qs,Murayama:2007ge,Roy:2007nz,Perez:2008ng,Komargodski:2008ax,Cho:2008fr,Asano:2008qc,Craig:2009rk,Evans:2012uf,Craig:2013wga,Knapen:2013zla,Craig:2008vs,Shirai:2008qt,Endo:2009uj,Craig:2009tz,Ding:2013pya}.
One of the interesting mechanisms in the hidden sector is conformal
sequestering, which can raise the mass of the gravitino relative to
the dangerous Planck-scale induced soft masses \cite{Craig:2008vs,Shirai:2008qt,Endo:2009uj,Craig:2009tz}.
So the lightest neutralino can be the LSP and DM candidate%
\footnote{Interestingly, the same mechanism can be used to solve the $\mu/B_{\mu}$-problem
in GMSB models \cite{Murayama:2007ge,Roy:2007nz,Cho:2008fr,Craig:2013wga,Knapen:2013zla}
or to construct focus point SUSY \cite{Ding:2013pya}.%
}. To illustrate these conformal sequestered models, we assume that
a strongly coupled hidden sector is approximately in a conformal window
$[M_{1},M_{2}]$, where $M_{2}$ is the scale at which the conformality
starts and $M_{\mathrm{1}}$ is the scale at which the conformality is broken.
Namely, $M_{EW}<M_{1}<M_{2}<M_{PL}$. In the conformal window, the
RGE runnings are dominated by the strongly coupled hidden sector.
As long as the fixed point is stable, the coupling constants flow
to their infrared fixed-point values by power laws. Below the conformal
window, one has 
\begin{equation}
b_{i,j}^{0}=\left(\frac{M_{1}}{M_{2}}\right)^{\beta_{S^{\dagger}S}}b_{i,j}=Z_{S^{\dagger}S}(M_{1})b_{i,j}.
\end{equation}
Here $Z_{S^{\dagger}S}(\mu)$ comes from one particle irreducible
(1PI) diagrams in the hidden sector deducting the wavefunction renormalization
factors. $\beta_{S^{\dagger}S}$ is the anomalous dimension of $S^{\dagger}S$.
Explicit models in the hidden sector have been discussed in \cite{Roy:2007nz,Shirai:2008qt,Craig:2009rk}
to demonstrate this conformal mechanism. If $\beta_{S^{\dagger}S}>0$,
$Z_{S^{\dagger}S}(M_{1})$ can offer a power suppressed factor which
is helpful to solve the flavor violation problem. Unfortunately, the
exact value of $\beta_{S^{\dagger}S}$ cannot be calculated in a perturbative
way. We simply assume that $b_{i,j}^{0}$ is small enough to be consistent
with the constraints on the flavor violation. So even if $m_{3/2}\sim\mathcal{O}(1\,\mathrm{TeV)}$,
the dangerous Planck-scale induced soft masses can be $m_{PL}^{\mathrm{soft}}\sim\sqrt{b_{ij}^{0}}m_{3/2}<1$
GeV. Gravitino will no longer be the LSP in GMSB models.

Besides the large anomalous dimension of $S^{\dagger}S$, the hidden
sector with sequestered SUSY breaking would also provide a significant
wavefunction renormalization factor $Z_{S}(\mu)$, which makes $\mathcal{L}_{\mathrm{eff}}=\int d^{4}\theta Z_{S}(\mu)S^{\dagger}S$
canonically normalized. $Z_{S}(\mu)$ can be absorbed into the redefinitions
of the couplings. For example, the coupling $\kappa$ in the superpotential
$W=\kappa S\Phi\bar{\Phi}$ becomes very small below the conformal
window as 
\begin{equation}
\kappa^{0}=\left(\frac{M_{1}}{M_{2}}\right)^{\frac{\gamma_{S}}{2}}\kappa=Z_{S}^{-\frac{1}{2}}(M_{1})\kappa.
\end{equation}
Here $\gamma_{S}$ is the anomalous dimension of $S$ at the conformal
fixed point. Since $S$ is a singlet, $\gamma_{S}=3R(S)/2-1$ with
$R(S)$ being the $R$ charge of $S$. The unitarity bound of the
superconformal algebra requires $R(S)>2/3$, which leads to $\gamma_{S}>1$
\cite{Dobrev:1985qv}. So the wavefunction renormalization always
offers a power suppressed factor to $\kappa$. Below the conformal window,
the superpotential is $W=\kappa^{0}S\Phi\bar{\Phi}$.

Finally we pay attention to the first term of Eq.(\ref{eq:kalher}), 
which is mediated by the gauge interaction. Since the superpotential
$W=\kappa^{0}S\Phi\bar{\Phi}$ contributes to the coefficient $c_{i}$,
$c_{i}$ must receive the $\gamma_{S}$ effect from anomalous dimension
of $S$. It is interesting to discuss whether this term will further
get a large correction from the anomalous dimension of $S^{\dagger}S$:

Case I: The messenger scale $M_{\mathrm{mess}}$ is below the conformal
window, namely $M_{EW}<M_{\mathrm{mess}}<M_{1}<M_{2}<M_{PL}$. After
the messengers fields are integrated out, the hidden sector is out
of the conformal window. Thus the coefficients $c_{i}$ do not receive the effect from the anomalous dimension $\beta_{S^{\dagger}S}$
\cite{Shirai:2008qt,Endo:2009uj}. Below the messenger scale, RGE
runnings, which are dominated by the traditional MSSM ones, allow
us to predict the entire MSSM spectrum at the electroweak scale. In
this case, the $\mu/B_{\mu}$-problem can be solved by introducing
some Planck-scale induced operators \cite{Shirai:2008qt}.

Case II: The messenger scale $M_{\mathrm{mess}}$ is within the conformal
window, namely $M_{EW}<M_{1}<M_{\mathrm{mess}}<M_{2}<M_{PL}$. After
the messengers fields are integrated out, the hidden sector is still
strongly coupled. Even the visible sector and hidden sector are coupled
through higher dimensional operators, the coefficient $c_{i}^{\mathrm{}}$
could be renormalized dominantly by the hidden sector. From the scale
$M_{\mathrm{mess}}$ to the scale $M_{1}$, $c_{i}$ will further
receive a damping factor. Below the scale $M_{1}$, all coefficients
run to the electoweak scale according to the usual MSSM RGEs. So in
this case the soft masses of sfermions will be further suppressed
by the large anomalous dimension of $S^{\dagger}S$ \cite{Craig:2008vs,Craig:2009tz}.
In order to make neutralino the LSP, the lightest stau should be heavier
than the lightest neutralino. This constraint in Case II is stronger
than that in Case I, since the stau mass in Case II will be further
suppressed. After the run I of LHC, a Higgs boson with mass around
125 GeV has been found but no SUSY particles have been detected. The
stop sector should provide a large loop contribution to raise Higgs
mass. Even assuming a non-vanishing $A_{t}-$term at the messenger
scale, stop mass would be heavier than 500 GeV to get a 125 GeV Higgs
\cite{Evans2013}. For the Case II, due to the suppression coming
from the anomalous dimension of $S^{\dagger}S$, it is hard to obtain
such heavy sfermions. A heavy stop may be realized if RGEs are assumed
to run for a long time. But this requirement asks for a high scale
$M_{1}$, which would weaken the suppression of the dangerous Plank-scale
induced operators. Thus, the Case II is not suggested by the current
LHC data. In the next section, we will discuss more phenomenologies
of the Case I.

\section{mass spectrum and neutralino dark matter}

\begin{figure}
\includegraphics[scale=0.6]{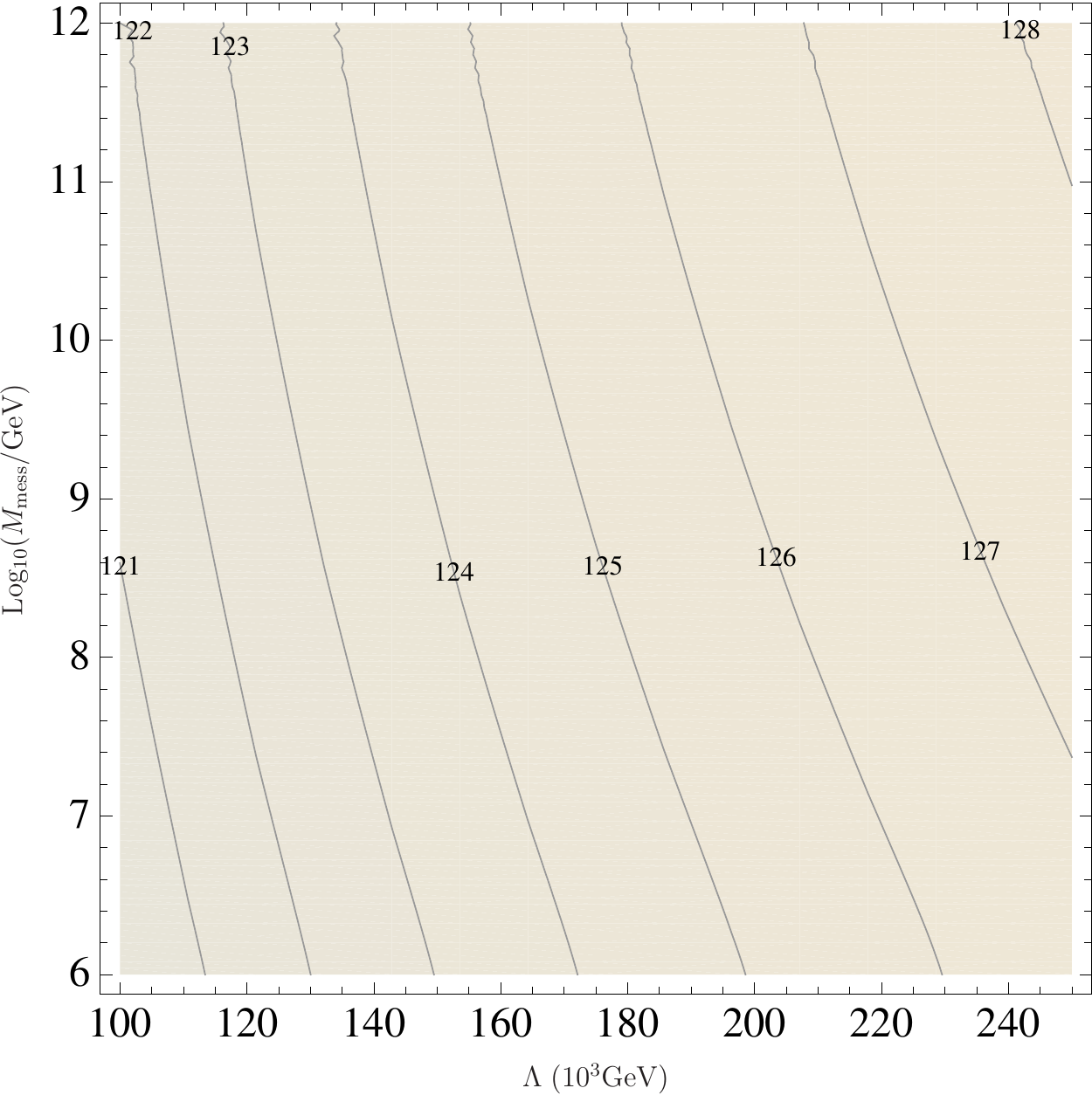} ~~~~\includegraphics[scale=0.6]{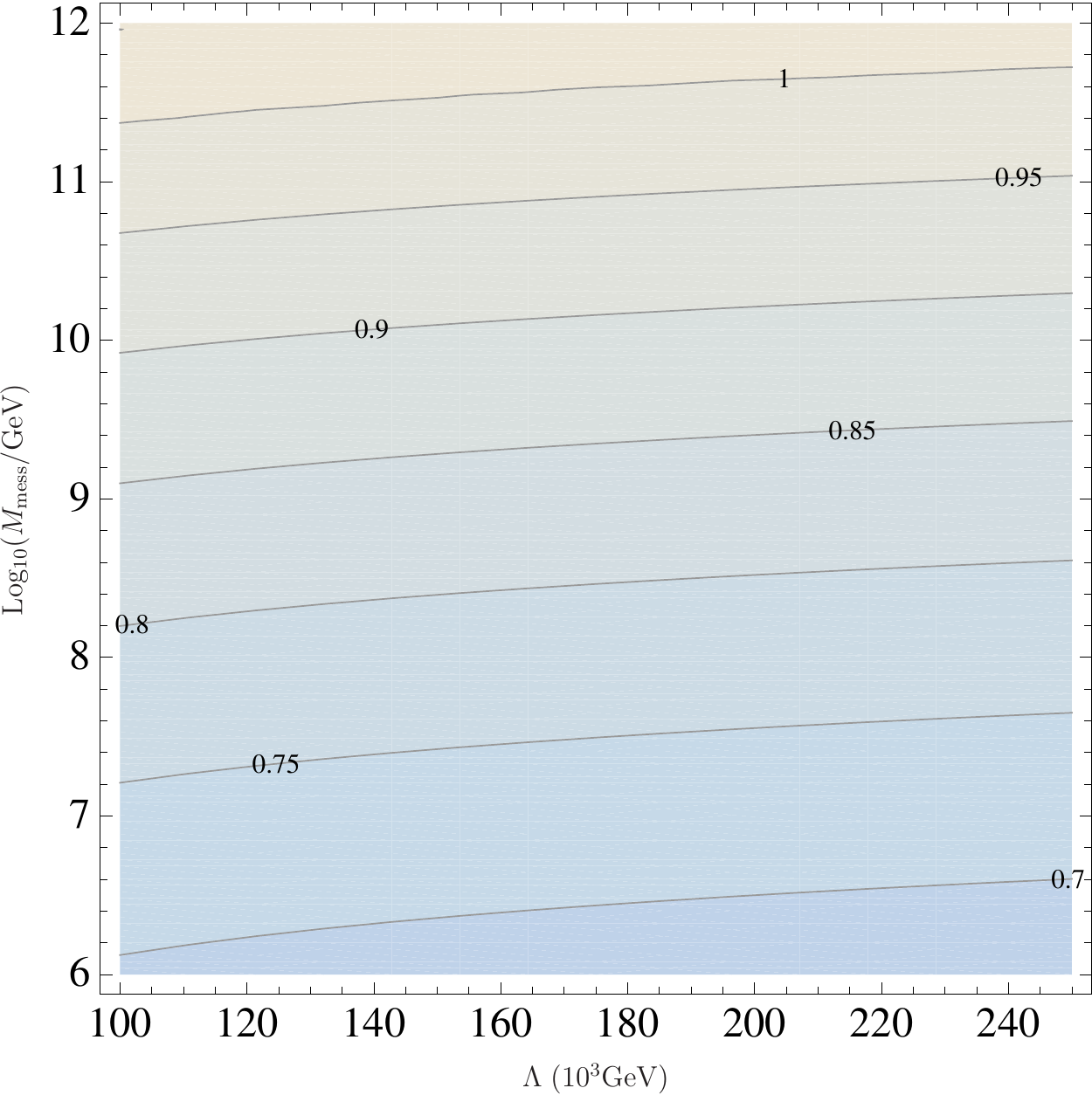}\caption{\label{fig:1}Contour plots of $m_{h}$ (left) and $m_{\tilde{{\tau}}_{1}}/m_{\tilde{{\chi}}_{1}^{0}}$
(right) in the $M_{\mathrm{{mess}}}$ vs. $\Lambda$ plane with $\tan\beta=10$.}
\end{figure}

In this section, we discuss MSSM mass spectrum and neutralino DM in
GMSB models with sequestered SUSY breaking. The gravitino mass is
fixed to be 1 TeV. We first study minimal GMSB model with $A=0$ at
the input scale. Then we move forward to an extension with non-vanishing
$A-$terms at the messenger scale.

\subsection{Minimal GMSB model with sequestered SUSY breaking}

In this model, the superpotential is 
\begin{equation}
W=\kappa S\Phi_{i}\bar{\Phi}_{i}.
\end{equation}
Here the messengers $\Phi_{i}$, $\bar{\Phi}_{i}$ fill out either
antisymmetric tensor $10+\overline{10}$ or fundamental $5+\bar{{5}}$
representation of $SU(5)$.
Below the conformal window, the conformal sequestered hidden sector
will lead to a very small coupling $\kappa^{0}$ in the superpotential,
which can be absorbed into the definition of mass parameter $\Lambda$
as $\Lambda=\frac{\kappa^{0}F_{S}}{M_{\mathrm{mess}}}$. This small
coupling $\kappa^{0}$ guarantees $\Lambda\sim\mathcal{O}(10^{5}\,\mathrm{GeV)}$
even when the gravitino mass is fixed to be 1 TeV. For the discussion
of phenomenologies, there are six input parameters as 
\begin{equation}
\left\{ \tan\beta,\;\mathrm{sign}(\mu),\; M_{\mathrm{mess}},\;\Lambda,\; n_{5},\; n_{10}\right\} .
\end{equation}
To perform a comprehensive analysis of our models, including spectrum
calculation and DM studies, we use the code\textsf{ toolbox1.2.2}
\cite{oai:arXiv.org:1109.5147}, which is compiled with \textsf{SARAH3.3.0},
\textsf{SPheno3.2.2} and \textsf{micrOMEGAs2.4.5}. The code\textsf{
SARAH} \cite{oai:arXiv.org:0909.2863,oai:arXiv.org:1002.0840,staub}
is used to create a \textsf{SPheno} version of our models with the
soft masses at the messenger scale. The mass spectrum at electroweak
scale is calculated by the code \textsf{SPheno} \cite{oai:arXiv.org:hep-ph/0301101,oai:arXiv.org:1104.1573}
with MSSM RGEs and the DM information is obtained by the code \textsf{micrOMEGAs
}\cite{oai:arXiv.org:0803.2360} %
\footnote{We calculate the mass of the Higgs boson at two-loop level. Recently,
some three-loop corrections have been discussed in \cite{Feng:2013tvd,Buchmueller:2013psa}.%
}. In our studies, $\mathrm{sign}(\mu)=+1$, $n_{5}=1$ and $n_{10}=1$
are fixed. We first scan the parameters $\Lambda$ and $M_{\mathrm{{mess}}}$
by assuming $\tan\beta=10$. Contour plots of $m_{h}$ in the $M_{\mathrm{{mess}}}$
vs. $\Lambda$ plane are shown in the left of Fig.(\ref{fig:1}).
For a fixed mass parameter $\Lambda$, the Higgs boson would be heavier
if the messenger scale is higher. Though $A_{t}=0$ at the messenger
scale, the $y_{t}M_{3}$ term in the RGE ensures that $A_{t}$ will
not vanish at the electroweak scale. RGE runnings also lift the stop
mass. A high-scale gauge mediation helps to obtain sufficiently large
absolute value of $A_{t}-$term and heavy stops at the electroweak
scale, which are preferred by a 125 GeV Higgs boson. In the right
of Fig.(\ref{fig:1}), we show the ratio of the lightest stau mass
to the lightest neutralino mass in the $M_{\mathrm{{mess}}}$ vs.
$\Lambda$ plane. In most of the parameter space, the LSP is the lightest
stau particle. A neutralino LSP can only be achieved when the messenger
scale $M_{\mathrm{mess}}$ is higher than $4\times10^{11}$ GeV.

\begin{figure}
\includegraphics[scale=0.6]{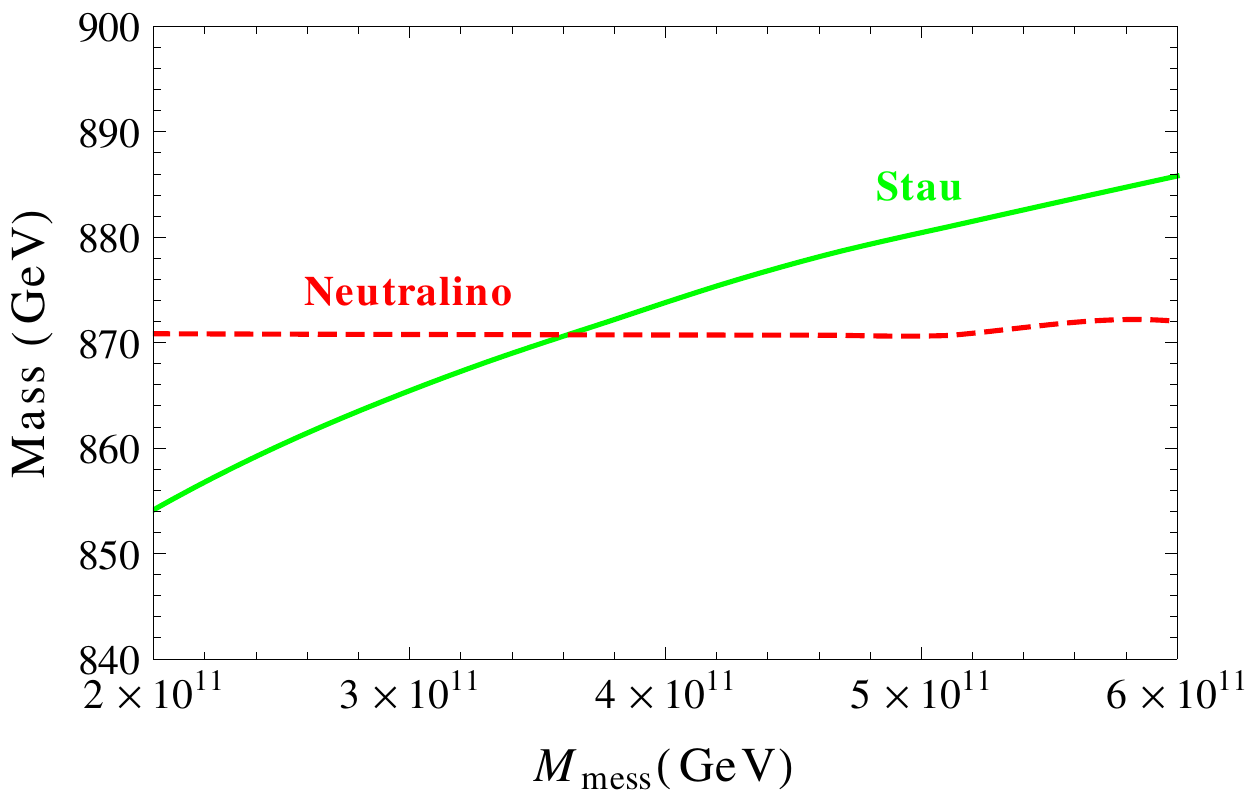} ~~~~\includegraphics[scale=0.6]{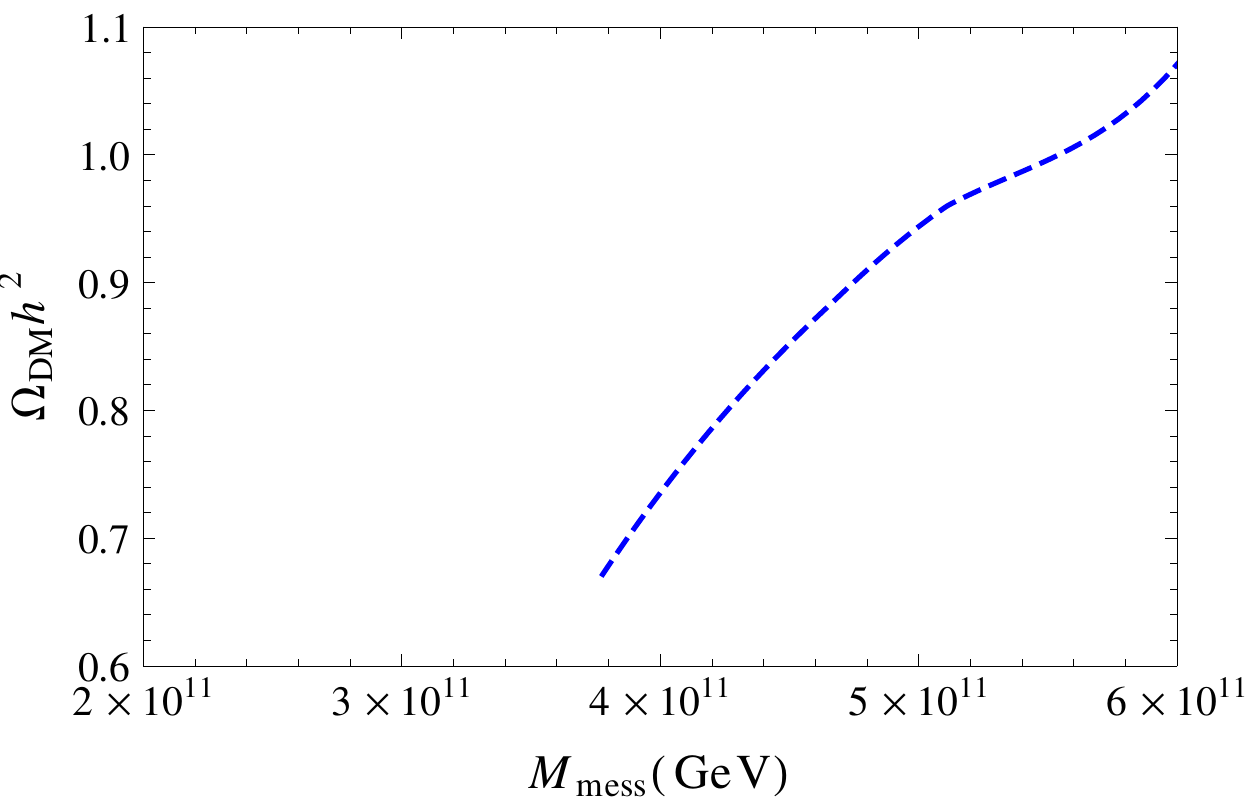}\caption{\label{fig:2}(color online) $\Lambda=1.6\times10^{5}$ GeV and $\tan\beta=10$.
Left: $m_{\tilde{}{\tau}_{1}}$ (green solid line) and $m_{\tilde{{\chi}}_{1}^{0}}$
(red dashed line) depend on the messenger scale $M_{\mathrm{{mess}}}$. Right:
The relic density $ $$\Omega\mathrm{{h}}^{2}$ depends on the messenger
scale $M_{\mathrm{{mess}}}$.}
\end{figure}

In Fig.(\ref{fig:2}), $\Lambda=1.6\times10^{5}$ GeV is fixed in
order to be consistent with a 125 GeV Higgs boson. In the left, we
show how the lightest stau mass $m_{\tilde{}{\tau}_{1}}$ and the
lightest neutralino mass $m_{\tilde{{\chi}}_{1}^{0}}$ depend on the
messenger scale $M_{\mathrm{{mess}}}$. In this case $\tilde{{\chi}}_{1}^{0}$
is purely bino-like and its mass is not sensitive to the messenger
scale $M_{\mathrm{{mess}}}$. Due to RGE running, $m_{\tilde{}{\tau}_{1}}$
becomes heavier for a higher messenger scale $M_{\mathrm{{mess}}}$.
When $M_{\mathrm{{mess}}}$ is larger than $3.6\times10^{11}$ GeV,
the LSP is $\tilde{{\chi}}_{1}^{0}$ and this model has a good DM
candidate with mass around 870 GeV. In the right, the DM relic density
$\Omega\mathrm{h}^{2}$ has been calculated by the code \textsf{micrOMEGAs}.
When the LSP is $\tilde{{\chi}}_{1}^{0}$, its relic density is always
larger than $0.6$, which is not consistent with the WMAP experimental
result $\Omega\mathrm{h}^{2}=0.1138\pm0.0045$ \cite{Bennett:2012zja}.
In this case, $\tilde{\tau}_{1}$ and $\tilde{{\chi}}_{1}^{0}$ are
degenerate and the coannihilation effect has been involved to make
predictions of relic density. Since the LSP is around 870 GeV, all
other SUSY particles should be heavier than 870 GeV. Because the exchanged
SUSY particles are so heavy, the cross section $\left\langle \sigma_{\mathrm{an}}v\right\rangle $
is not large enough even including the coannihilation effect. That
is why we get too large DM relic density in this model. We have varied
the value of $\tan\beta$ in this model. But the main features of
Fig.(\ref{fig:1}) and Fig.(\ref{fig:2}) do not change. DM candidate
is purely bino-like with a relatively large mass. It is well-known that
the observed relic abundance requires the mass of purely bino-like
DM to be less than 200 GeV for thermal production \cite{ArkaniHamed:2006mb}.
Even including coannihilation effects, purely bino-like DM cannot
be too heavy \cite{Ellis:1998kh}. So generally speaking,
the neutralino DM with correct relic density is hard to be achieved
in this model.

\subsection{An extension model with non-vanishing $A-$terms}

Minimal GMSB model can be extended with non-vanishing $A-$terms at
the messenger scale. In \cite{Kang:2012ra,Craig:2012xp,Albaid:2012qk,Zheng:2013lga,Ding:2013pya,Byakti:2013ti,Evans2013},
new Yukawa couplings between the Higgs sector and messengers are introduced
to generate one-loop $A$-terms at $M_{\mathrm{{mess}}}$ scale without
flavor problems. So in this subsection, we add a new term in the superpotential
as 
\begin{equation}
\triangle W=\lambda_{u}H_{u}\Phi_{i}\Phi_{S}.\label{eq:lambda_u}
\end{equation}
Here we introduce a new singlet $\Phi_{S}$ as another messenger field. $\Phi_{i}$
are all the fields taking the $(\mathbf{1},\mathbf{2},-1/2)$ representation
in the $5+\bar{{5}}$ messenger fields. Eq.(\ref{eq:lambda_u}) would
lead to a non-vanishing $A_{t}$ at the messenger scale. Since the
singlet $S$ is the only SUSY-breaking source, the $A/m_{H_{u}}^{2}$-problem
is not large \cite{Craig:2012xp}. Here we do not introduce new Yukawa
couplings between $H_{d}$ and the messenger fields. So there is no
$\mu/B_{\mu}$-problem. In this GMSB model with sequestered SUSY breaking,
the $\mu$-term can be generated by some Planck-scale induced operators
\cite{Shirai:2008qt}. Compared to the mass spectrum in minimal GMSB
model, Eq.(\ref{eq:lambda_u}) results in extra contributions of $A_{t}$,
$m_{H_{u}}^{2}$, $m_{Q}^{2}$ and $m_{U}^{2}$ at the input scale
as \cite{Craig:2012xp} 
\begin{equation}
\begin{cases}
A_{t} & =-\frac{n_{5}\lambda_{u}^{2}}{16\pi^{2}}\Lambda,\\
m_{H_{u}}^{2} & =-\frac{n_{5}\lambda_{u}^{2}}{48\pi^{2}}h\left(\frac{\Lambda}{M_{\mathrm{mess}}}\right)\left(\frac{\Lambda}{M_{\mathrm{mess}}}\right)^{2}\Lambda^{2}+\frac{(3+n_{5})\lambda_{u}^{4}-(3g_{1}^{2}/5+3g_{2}^{2})\lambda_{u}^{2}}{256\pi^{4}}n_{5}\Lambda^{2},\\
m_{Q}^{2} & =-\frac{n_{5}y_{t}^{2}\lambda_{u}^{2}}{256\pi^{4}}\Lambda^{2},\\
m_{U}^{2} & =-\frac{n_{5}y_{t}^{2}\lambda_{u}^{2}}{128\pi^{4}}\Lambda^{2}.
\end{cases}\label{eq:correction_mass}
\end{equation}
Here the function $h(x)\approx1+4x^{2}/5$. If the messenger scale
$M_{\mathrm{mess}}\sim\mathcal{O}(10^{5}\,\mathrm{GeV)}$, the first
term of $m_{H_{u}}^{2}$ in Eq.(\ref{eq:correction_mass}) is important
to realize the electoweak symmetry breaking (EWSB). When the messenger
scale $M_{\mathrm{mess}}$ is large, this term can be neglected due
to the $M_{\mathrm{mess}}$-suppression. Instead, the top Yukawa $y_{t}$
contribution in the RGEs could cause $m_{H_{u}}^{2}$ to run negative
at the electroweak scale, helping to achieve EWSB.

\begin{figure}
\includegraphics[scale=0.6]{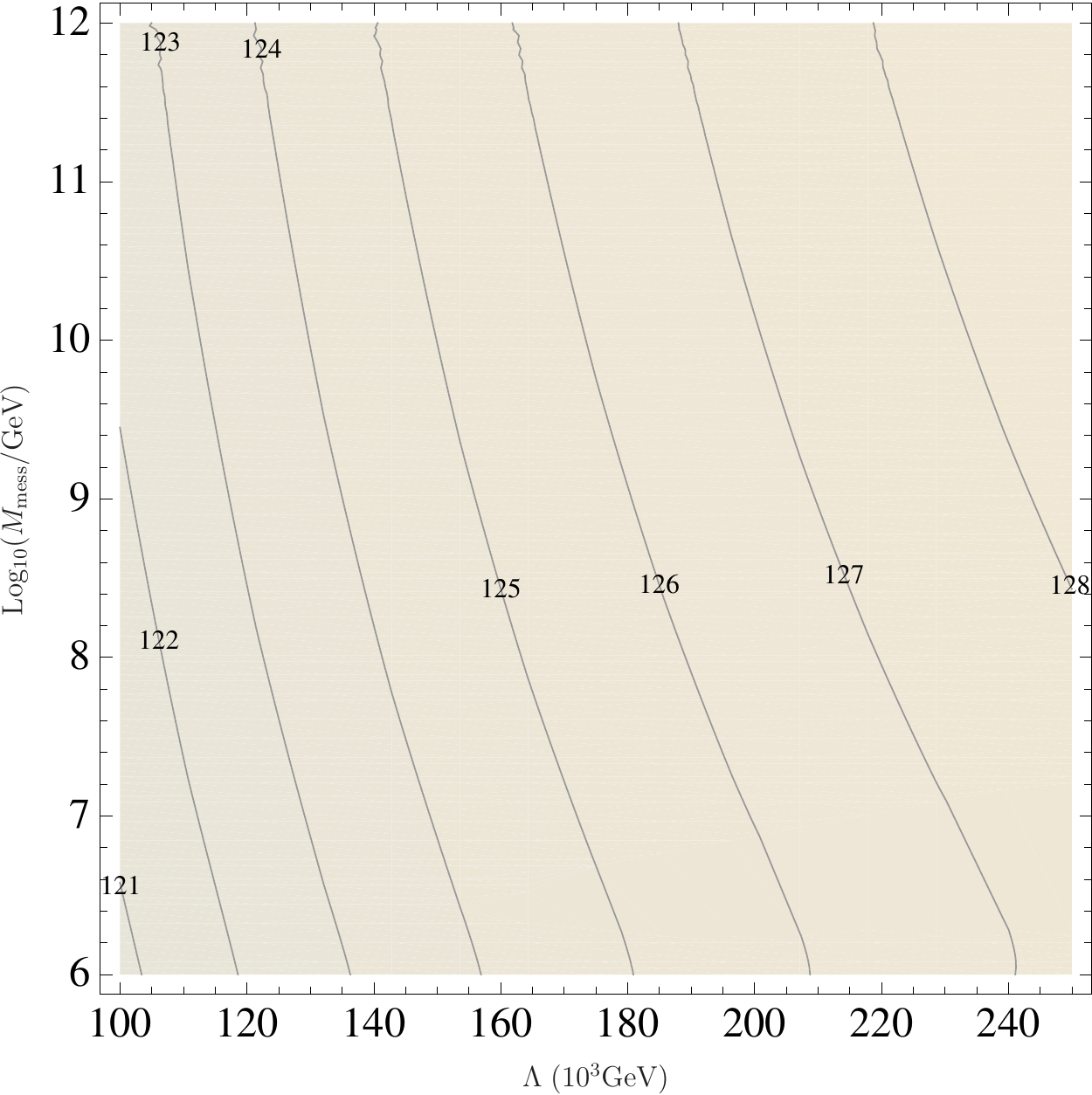} ~~~~\includegraphics[scale=0.6]{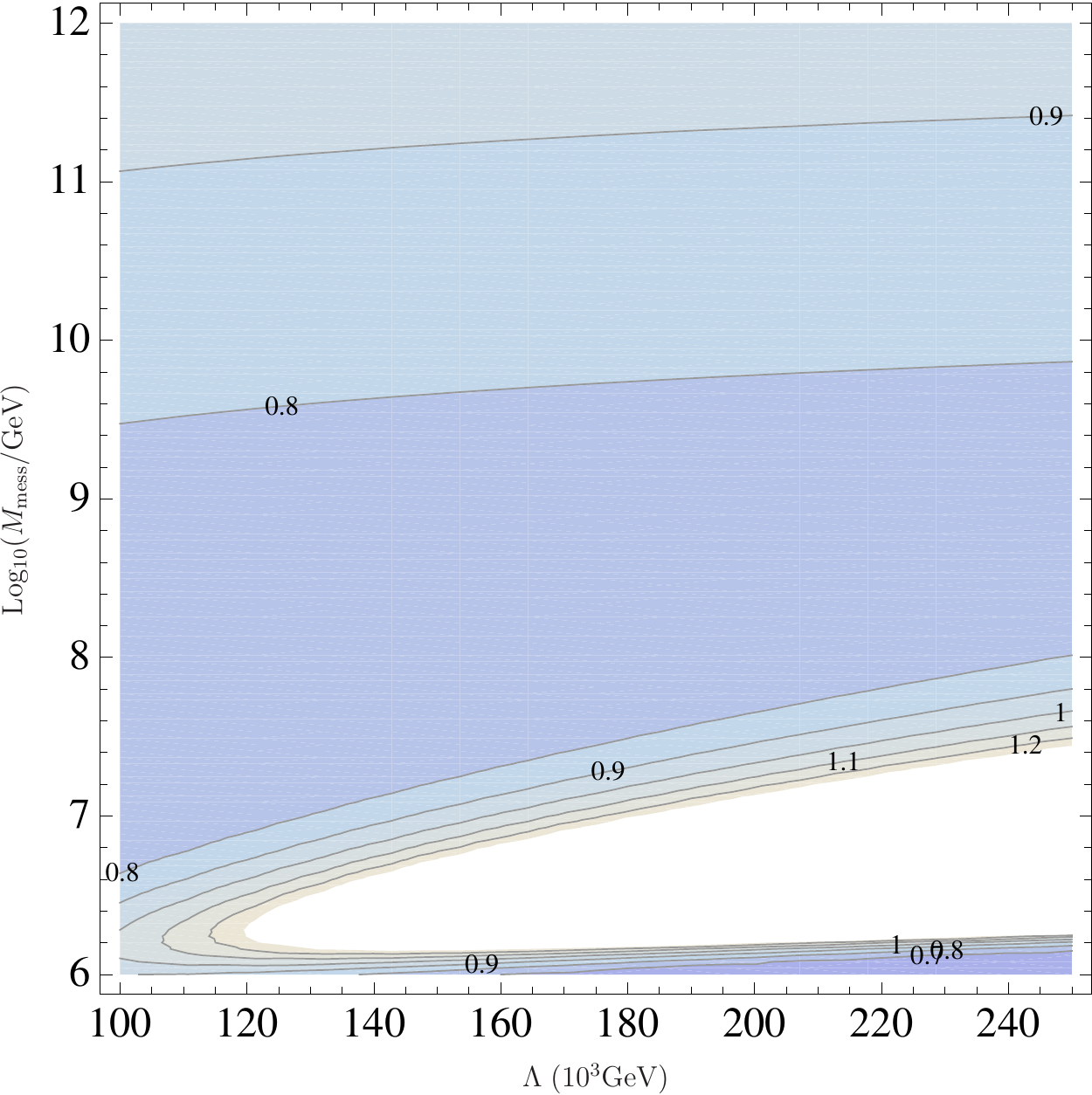}\caption{\label{fig:3}Contour plots of $m_{h}$ (left) and $m_{\tilde{{\tau}}_{1}}/m_{\tilde{{\chi}}_{1}^{0}}$
(right) in the $M_{\mathrm{{mess}}}$ vs. $\Lambda$ plane with $\tan\beta=10$
and $\lambda_{u}=1$. In the whole blank area of right figure, $m_{\tilde{{\tau}}_{1}}/m_{\tilde{{\chi}}_{1}^{0}}>1$.
Since $m_{\tilde{{\tau}}_{1}}/m_{\tilde{{\chi}}_{1}^{0}}$ is very
sensitive to the choice of $\Lambda$ and $M_{\mathrm{mess}}$ in
this area, the exact values are difficult to be shown in this contour.}
\end{figure}

So in this model, there are seven input parameters as 
\begin{equation}
\left\{ \tan\beta,\;\mathrm{sign}(\mu),\; M_{\mathrm{mess}},\;\Lambda,\;\lambda_{u},\; n_{5},\; n_{10}\right\} .
\end{equation}
$\lambda_{u}$ is not suppressed by the sequestered SUSY breaking
sector since it is not directly coupled to the hidden sector $S$.
Thus $\lambda_{u}\sim\mathcal{O}(1)$. Contour plots of $m_{h}$ and
$m_{\tilde{{\tau}}_{1}}/m_{\tilde{{\chi}}_{1}^{0}}$ in the $M_{\mathrm{{mess}}}$
vs. $\Lambda$ plane are shown in Fig.(\ref{fig:3}) when $\tan\beta=10$
and $\lambda_{u}=1$ are assumed. By comparing the left figures between
Fig.(\ref{fig:1}) and Fig.(\ref{fig:3}), the Higgs boson with mass
around 125 GeV is easier to be obtained with non-vanishing $A$-term.
In the right of Fig.(\ref{fig:3}), we show the ratio of the lightest
stau mass to the lightest neutralino mass in the $M_{\mathrm{{mess}}}$
vs. $\Lambda$ plane. A neutralino LSP as well as a 125 GeV Higgs
can be achieved in a large parameter space with $10^{6}\,\mathrm{GeV}<M_{\mathrm{mess}}<10^{7}\,\mathrm{GeV}$,
as shown in the blank area in the right of Fig.(\ref{fig:3}). We
should like to focus on neutralino DM in this parameter area.

\begin{figure}
\includegraphics[scale=0.6]{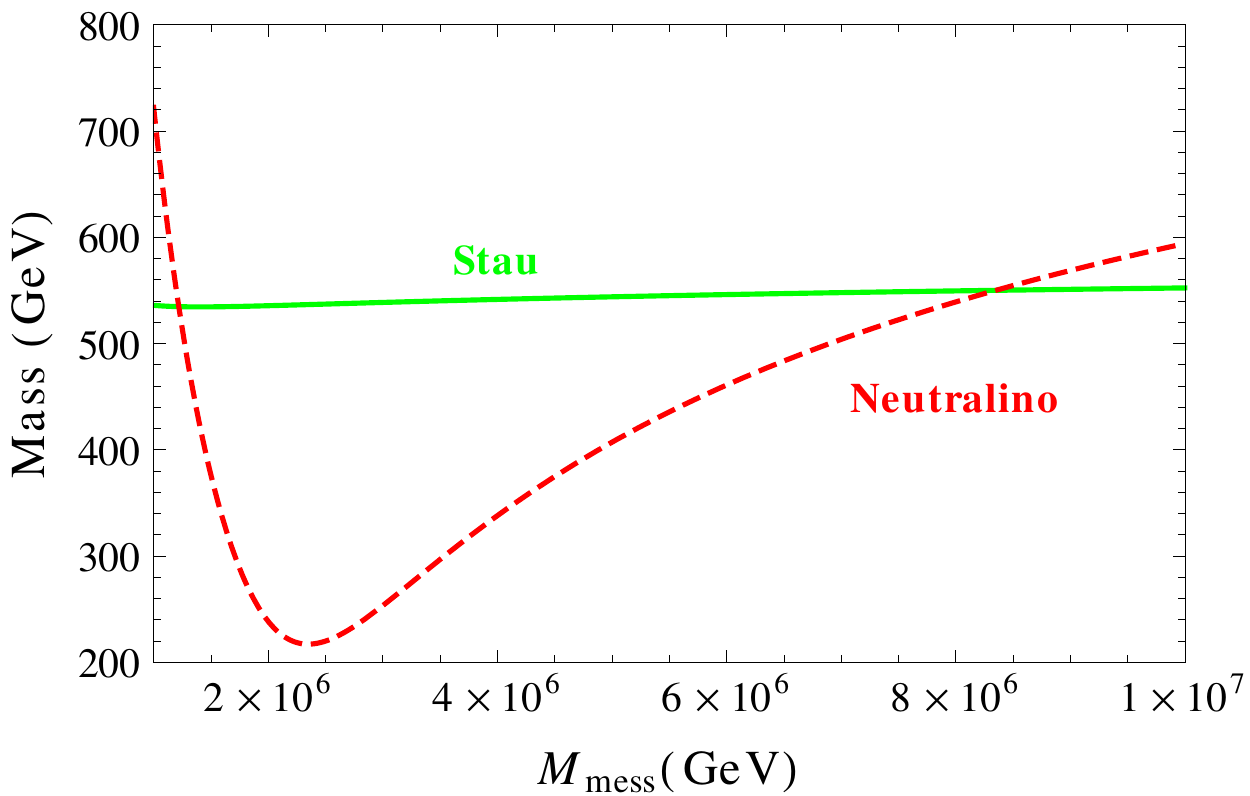} ~~~~\includegraphics[scale=0.6]{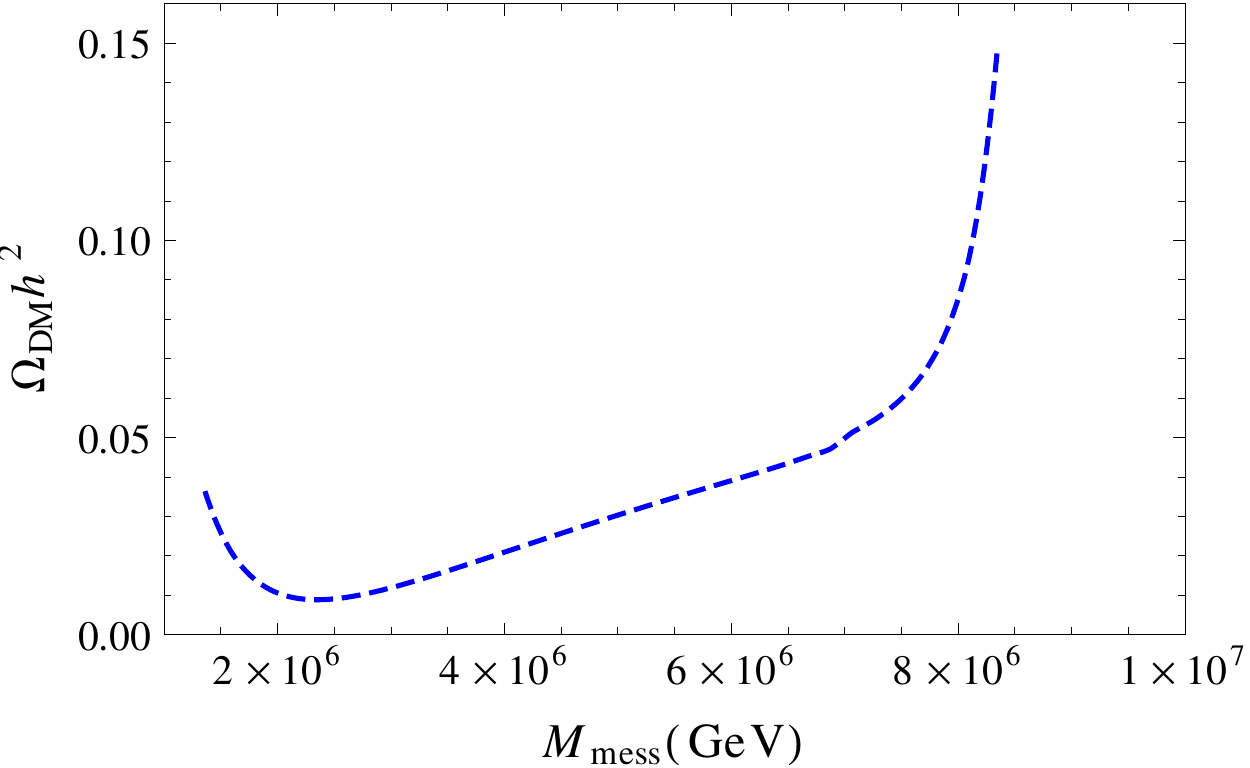}\caption{\label{fig:4}(color online) $\Lambda=1.5\times10^{5}$ GeV, $\tan\beta=10$ and
$\lambda_{u}=1$. Left: $m_{\tilde{}{\tau}_{1}}$ (green solid line) and $m_{\tilde{{\chi}}_{1}^{0}}$
(red dashed line) depend on the messenger scale $M_{\mathrm{{mess}}}$. Right:
The relic density $ $$\Omega\mathrm{{h}}^{2}$ depends on the messenger
scale $M_{\mathrm{{mess}}}$. }
\end{figure}

In Fig.(\ref{fig:4}), $\Lambda=1.5\times10^{5}$ GeV is fixed in
order to be consistent with a 125 GeV Higgs boson. In the left, we
show how the lightest stau mass $m_{\tilde{}{\tau}_{1}}$ and the
lightest neutralino mass $m_{\tilde{{\chi}}_{1}^{0}}$ depend on the
messenger scale in the range $10^{6}\,\mathrm{GeV}<M_{\mathrm{mess}}<10^{7}\,\mathrm{GeV}$.
In this range, $m_{\tilde{}{\tau}_{1}}$ is almost independent of
the messenger scale and $\tilde{{\chi}}_{1}^{0}$ is actually a mixture
of bino and higgsino. $m_{\tilde{{\chi}}_{1}^{0}}$ is sensitive to
the messenger scale because $m_{\tilde{{\chi}}_{1}^{0}}$ is dominated
by the value of $\mu$-term, which depends on $M_{\mathrm{{mess}}}$.
The exact value of $\mu$-term is determined by the correct EWSB.
Due to the $\lambda_{u}$ corrections of $m_{H_{u}}^{2}$ in Eq.(\ref{eq:correction_mass}),
EWSB in this model is quite different from that in the minimal GMSB
model. In the range $10^{6}\,\mathrm{GeV}<M_{\mathrm{mess}}<10^{7}\,\mathrm{GeV}$,
EWSB can be realized by two reasons. One is the negative $\Lambda/M_{\mathrm{mess}}$-suppressed
contribution of $m_{H_{u}}^{2}$ at input scale and the other is the
top Yukawa contribution in RGE running. In the range $1.5\times10^{6}\,\mathrm{GeV}<M_{\mathrm{mess}}<8\times10^{6}\,\mathrm{GeV}$,
the correct EWSB hints that $\mu$-term is less than 500 GeV, which
makes $\tilde{{\chi}}_{1}^{0}$ lighter than $\tilde{\tau}_{1}$.
As it is a bino-higgsino DM, the corresponding DM relic density $\Omega\mathrm{h}^{2}$
has been shown in the right of Fig.(\ref{fig:4}). We can have a neutralino
DM which is consistent with the WMAP experimental relic density result
$\Omega\mathrm{h}^{2}=0.1138\pm0.0045$ \cite{Bennett:2012zja}. Though
we fix $\Lambda=1.5\times10^{5}$ GeV in the above discussion, our
conclusion is general. A relatively small $\mu$-term can be obtained
in this model, which makes $\tilde{{\chi}}_{1}^{0}$ the LSP. So a
bino-higgsino DM with correct relic density can be achieved. On the
other hand, EWSB with a large $\tan\beta$ leads to the following
constraint at the electroweak scale, 
\begin{equation}
m_{Z}^{2}\approx-2(\mu^{2}+m_{H_{u}}^{2}).
\end{equation}
Since the value of $\mu$-term is relatively small in this model,
the cancellation between $\mu$ and $m_{H_{u}}$ is correspondingly
relatively small. There is a small fine-tuning to get the $Z$ boson
mass.

\begin{figure}
\includegraphics[scale=0.6]{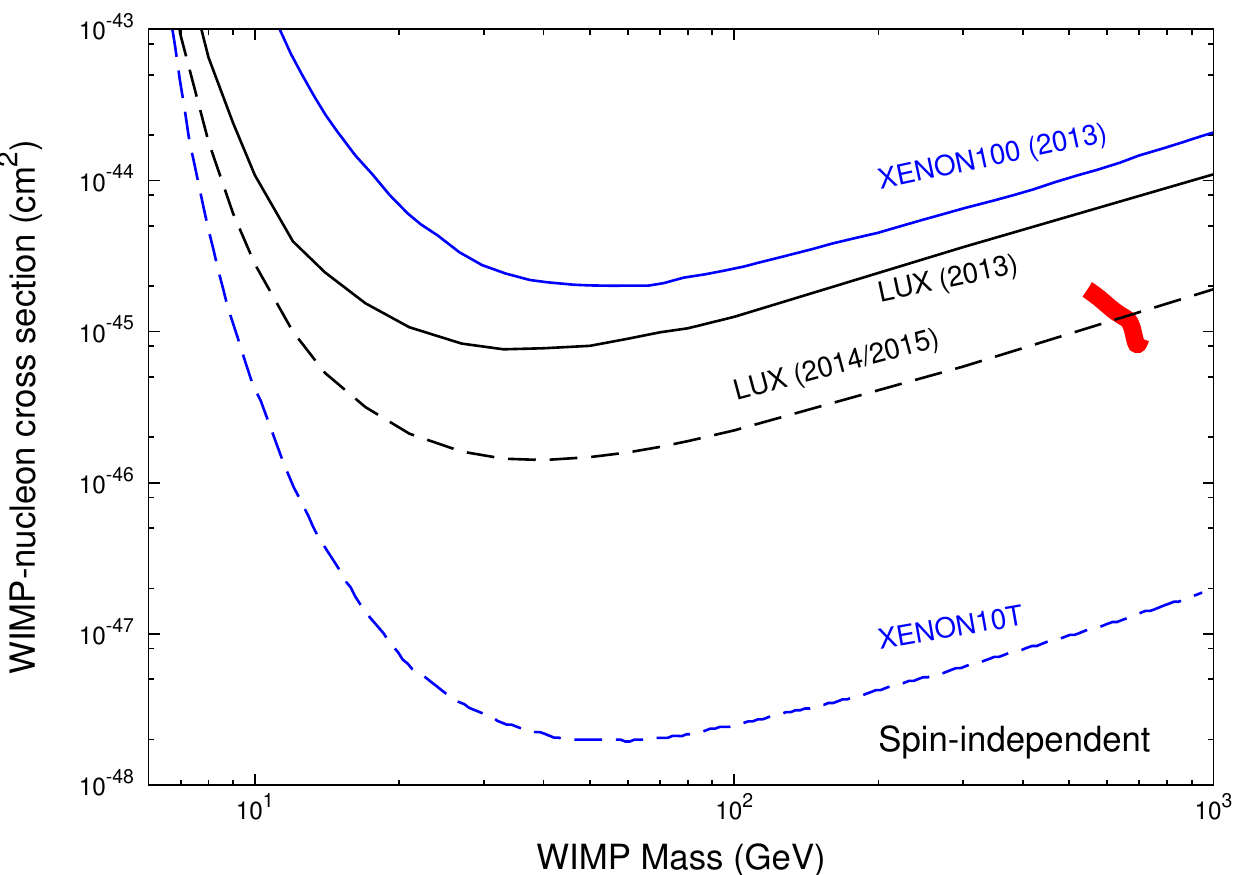} ~~~~\includegraphics[scale=0.6]{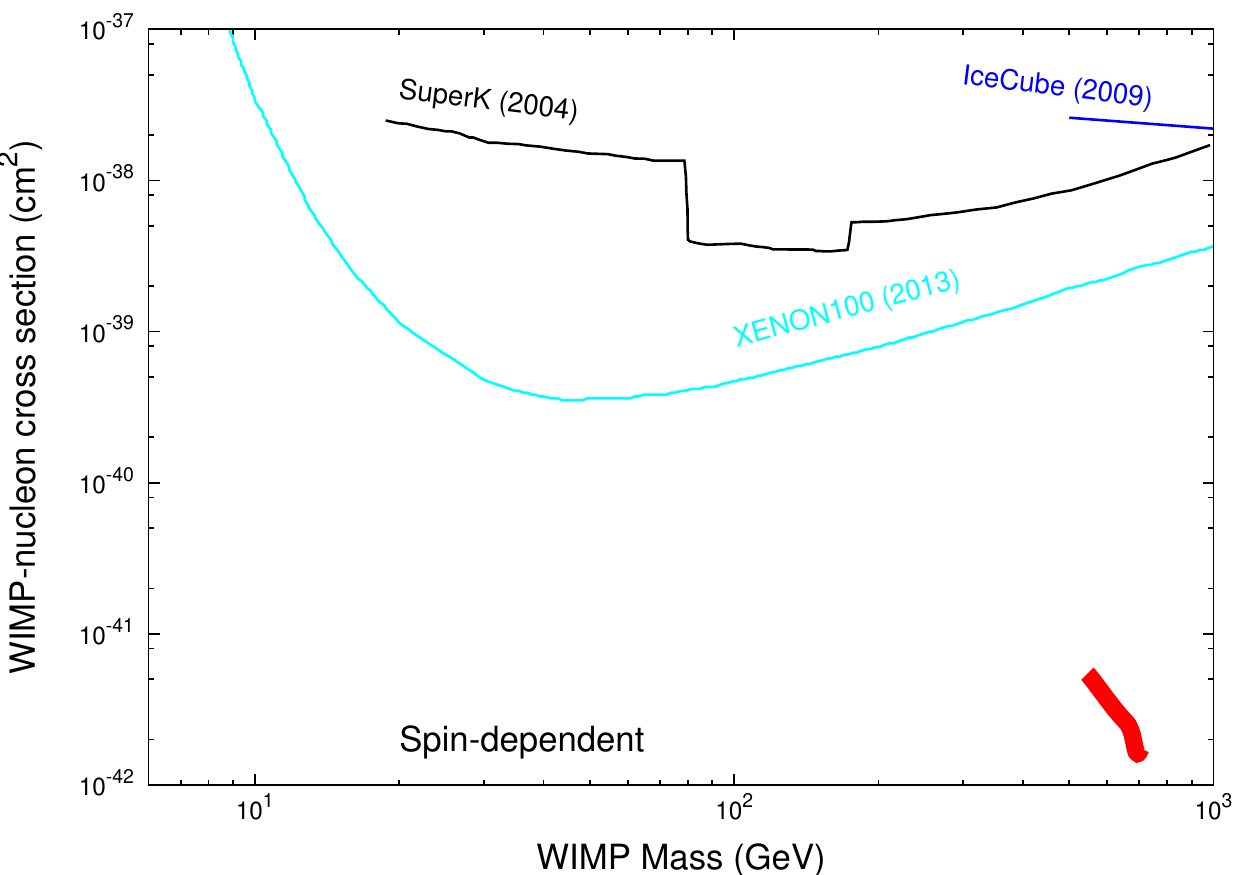}\caption{\label{fig:5} (color online) Our DM points are shown in the red region. For the spin-independent
cross section, plot $\sigma_{\mathrm{SI}}$ vs. $m_{\tilde{{\chi}}_{1}^{0}}$
is shown in the left, with the current bounds from LUX \cite{Akerib:2013tjd}
(solid black line), XENON100 \cite{Aprile:2012nq} (solid blue line)
and future reaches of LUX(2014/2015) \cite{LUX2015} (dashed black
line), XENON10T \cite{XENON10T} (dashed blue line). For the spin-dependent
cross section, plot $\sigma_{\mathrm{SD}}$ vs. $m_{\tilde{{\chi}}_{1}^{0}}$
is shown in the right, with the current bounds from SuperK \cite{Desai:2004pq}
(solid black line), IceCube \cite{Abbasi:2009uz} (solid blue line)
and XENON100 \cite{Aprile:2013doa} (solid cyan line).}
\end{figure}

Finally, we take into account the updated bounds of DM direct searches.
The current strictest bound of spin-independent cross section is recently
given by the LUX Collaboration \cite{Akerib:2013tjd}, who is the
first to break the $10^{-45}\;\mathrm{cm^{2}}$ cross section barrier
of DM spin-independent detection. We also consider the existing upper
limits of spin-dependent cross section. For this study, we scan the
the parameters in the $M_{\mathrm{{mess}}}$ vs. $\Lambda$ plane
and collect the points which have a Higgs boson with mass $123\,\mathrm{GeV}<m_{h}<127\,\mathrm{GeV}$
and a bino-higgsino DM with relic density $0.1<\Omega\mathrm{h}^{2}<0.12$.
The results of DM direct searches are shown in Fig.(\ref{fig:5}).
The left figure is devoted to the spin-independent cross section.
Our DM points are below the current experimental bounds, such as LUX
\cite{Akerib:2013tjd} and XENON100 \cite{Aprile:2012nq}. Interestingly,
based on the proposals of future experiments, our DM points can be
examined by future DM direct searches, such as LUX in 2015 \cite{LUX2015}
and XENON10T \cite{XENON10T}. For the spin-dependent cross section,
the results are shown in the right figure. Our DM points are far below
the existing experimental bounds. For both spin-independent detection and spin-dependent detection,  
the cross section will become relatively small if DM is relatively heavy.
That is because all other SUSY particles should be heavier than the LSP. 
DM with a relatively large mass will force overall sparticles to be relatively heavy. 

\begin{figure}
\includegraphics[scale=0.8]{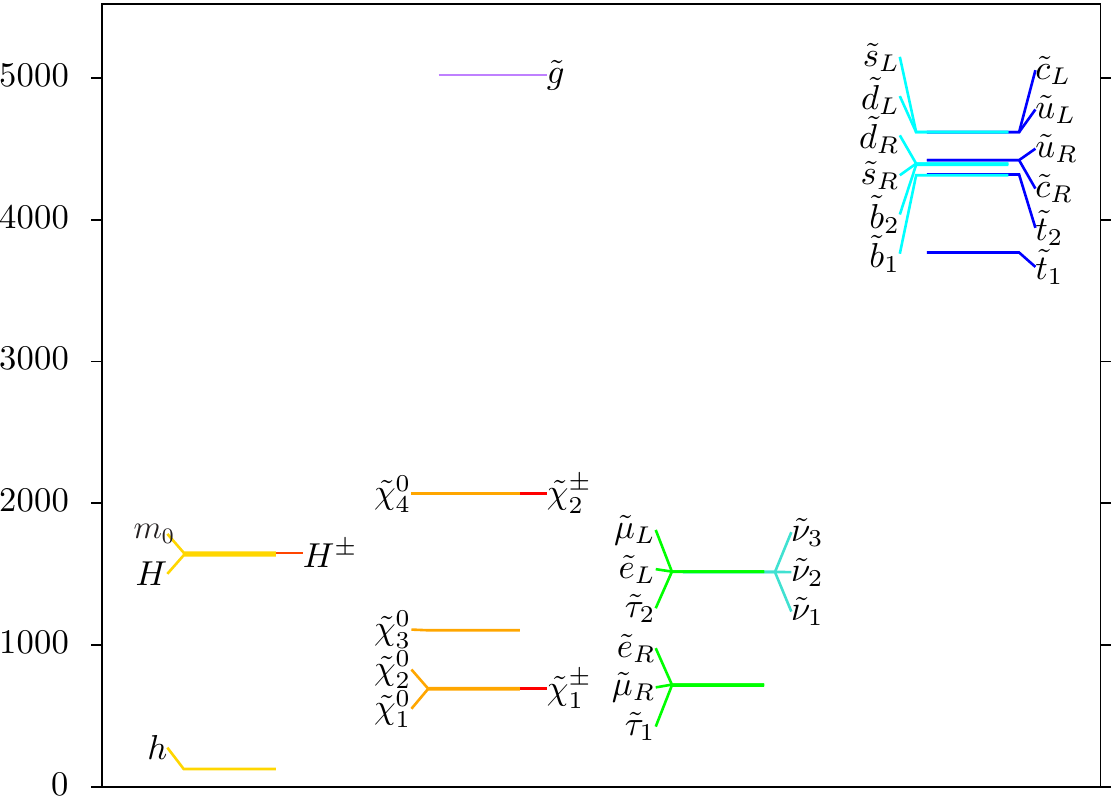}\caption{\label{fig:6} Mass spectrum of a benchmark point. In this case a
bino-higgsino DM with right relic density is predicted. }
\end{figure}

\section{Conclusion}

In this paper, we have studied the neutralino DM in gauge mediation
using the data after the run I of LHC and LUX. Neutralino can be the
DM candidate in GMSB models if the conformal sequestered mechanism
is introduced in the hidden sector. So the gravitino mass $m_{3/2}$
can be fixed to 1 TeV without introducing any flavor violation problem.
For the minimal GMSB model with sequestered SUSY breaking, the DM
candidate can be a purely bino-like neutralino. In this case it is hard to
achieve the correct relic density due to its relatively large mass.
So we move forward to extending the minimal GMSB model by adding new
Yukawa couplings between the messenger fields and the Higgs field
$H_{u}$. In this extension, this sequestered mechanism can predict
a good DM candidate as well as a 125 GeV Higgs boson. As an example,
the mass spectrum of one benchmark point is shown in Fig.(\ref{fig:6}),
which is corresponding to $m_{\tilde{{\chi}}_{1}^{0}}=688.4\,\mathrm{GeV}$ and $\Omega\mathrm{h}^{2}=0.108$.
The initial parameters are $\mathrm{sign}(\mu)=+1$, $n_{5}=1$, $n_{10}=1$,
$\tan\beta=10$, $\lambda_{u}=1$, $\Lambda=2\times10^{5}\,\mathrm{GeV}$
and $M_{\mathrm{mess}}=1.46\times10^{6}\,\mathrm{GeV}$. Thus for
this case, the coupling is 
\begin{equation}
\kappa^{0}\sim\frac{\Lambda M_{\mathrm{mess}}}{m_{3/2}M_{PL}}\sim\mathcal{O}(10^{-10}).
\end{equation}
This $\kappa^{0}$ can be simply realized, for example, by assuming
$M_{1}=2\times10^{6}\,\mathrm{GeV}$, $M_{2}=2\times10^{16}\,\mathrm{GeV}$
and $\gamma_{S}=2$. $\gamma_{S}=2$ can be achieved if the hidden
sector is $SP(3)\times SP(1)^{2}$ model. All our survived points
have some common features. Firstly, the light Higgs boson $h$ is
around 125 GeV and other Higgs bosons are heavy. So the Higgs sector
falls into the decoupling MSSM limit. The properties of the light
Higgs boson $h$ are similar to the predictions of the SM Higgs boson.
Secondly, the correct EWSB hints a relatively small $\mu$-term, which
makes the lightest neutralino lighter than the lightest stau. So a
bino-higgsino DM with correct relic density
can be achieved. The relatively small $\mu$-term results in a small
fine-tuning of obtaining the $Z$ boson mass. Finally, this bino-higgsino
DM can pass all the existing bounds of both spin-independent and spin-dependent
searches. Interestingly, the spin-independent cross section of our
DM points can be examined by further dark matter experiments, such
as LUX in 2015 and XENON10T. 
\begin{acknowledgments}
We would like to thank Qaisar Shafi, Ilia Gogoladze, Kai Wang, David
Shih, Florian Staub and Jared Evans for very valuable discussions
or comments. L.W. is supported by the DOE Grant No. DE-FG02-12ER41808.\bibliographystyle{h-physrev}
\bibliography{DM_125GeV}

\begin{thebibliography}{115}
\expandafter\ifx\csname natexlab\endcsname\relax\def\natexlab#1{#1}\fi
\expandafter\ifx\csname bibnamefont\endcsname\relax
  \def\bibnamefont#1{#1}\fi
\expandafter\ifx\csname bibfnamefont\endcsname\relax
  \def\bibfnamefont#1{#1}\fi
\expandafter\ifx\csname citenamefont\endcsname\relax
  \def\citenamefont#1{#1}\fi
\expandafter\ifx\csname url\endcsname\relax
  \def\url#1{\texttt{#1}}\fi
\expandafter\ifx\csname urlprefix\endcsname\relax\def\urlprefix{URL }\fi
\providecommand{\bibinfo}[2]{#2}
\providecommand{\eprint}[2][]{\url{#2}}

\bibitem[{\citenamefont{Dine and Fischler}(1982)}]{Dine:1981gu}
\bibinfo{author}{\bibfnamefont{M.}~\bibnamefont{Dine}} \bibnamefont{and}
  \bibinfo{author}{\bibfnamefont{W.}~\bibnamefont{Fischler}},
  \bibinfo{journal}{Phys.Lett.} \textbf{\bibinfo{volume}{B110}},
  \bibinfo{pages}{227} (\bibinfo{year}{1982}).

\bibitem[{\citenamefont{Dine et~al.}(1981)\citenamefont{Dine, Fischler, and
  Srednicki}}]{Dine:1981za}
\bibinfo{author}{\bibfnamefont{M.}~\bibnamefont{Dine}},
  \bibinfo{author}{\bibfnamefont{W.}~\bibnamefont{Fischler}}, \bibnamefont{and}
  \bibinfo{author}{\bibfnamefont{M.}~\bibnamefont{Srednicki}},
  \bibinfo{journal}{Nucl.Phys.} \textbf{\bibinfo{volume}{B189}},
  \bibinfo{pages}{575} (\bibinfo{year}{1981}).

\bibitem[{\citenamefont{Dimopoulos and Raby}(1981)}]{Dimopoulos:1981au}
\bibinfo{author}{\bibfnamefont{S.}~\bibnamefont{Dimopoulos}} \bibnamefont{and}
  \bibinfo{author}{\bibfnamefont{S.}~\bibnamefont{Raby}},
  \bibinfo{journal}{Nucl.Phys.} \textbf{\bibinfo{volume}{B192}},
  \bibinfo{pages}{353} (\bibinfo{year}{1981}).

\bibitem[{\citenamefont{Nappi and Ovrut}(1982)}]{Nappi:1982hm}
\bibinfo{author}{\bibfnamefont{C.~R.} \bibnamefont{Nappi}} \bibnamefont{and}
  \bibinfo{author}{\bibfnamefont{B.~A.} \bibnamefont{Ovrut}},
  \bibinfo{journal}{Phys.Lett.} \textbf{\bibinfo{volume}{B113}},
  \bibinfo{pages}{175} (\bibinfo{year}{1982}).

\bibitem[{\citenamefont{Alvarez-Gaume et~al.}(1982)\citenamefont{Alvarez-Gaume,
  Claudson, and Wise}}]{AlvarezGaume:1981wy}
\bibinfo{author}{\bibfnamefont{L.}~\bibnamefont{Alvarez-Gaume}},
  \bibinfo{author}{\bibfnamefont{M.}~\bibnamefont{Claudson}}, \bibnamefont{and}
  \bibinfo{author}{\bibfnamefont{M.~B.} \bibnamefont{Wise}},
  \bibinfo{journal}{Nucl.Phys.} \textbf{\bibinfo{volume}{B207}},
  \bibinfo{pages}{96} (\bibinfo{year}{1982}).

\bibitem[{\citenamefont{Dine and Nelson}(1993)}]{Dine:1993yw}
\bibinfo{author}{\bibfnamefont{M.}~\bibnamefont{Dine}} \bibnamefont{and}
  \bibinfo{author}{\bibfnamefont{A.~E.} \bibnamefont{Nelson}},
  \bibinfo{journal}{Phys.Rev.} \textbf{\bibinfo{volume}{D48}},
  \bibinfo{pages}{1277} (\bibinfo{year}{1993}), \eprint{hep-ph/9303230}.

\bibitem[{\citenamefont{Dine et~al.}(1993)\citenamefont{Dine, Leigh, and
  Kagan}}]{Dine:1993qm}
\bibinfo{author}{\bibfnamefont{M.}~\bibnamefont{Dine}},
  \bibinfo{author}{\bibfnamefont{R.~G.} \bibnamefont{Leigh}}, \bibnamefont{and}
  \bibinfo{author}{\bibfnamefont{A.}~\bibnamefont{Kagan}},
  \bibinfo{journal}{Phys.Rev.} \textbf{\bibinfo{volume}{D48}},
  \bibinfo{pages}{2214} (\bibinfo{year}{1993}), \eprint{hep-ph/9303296}.

\bibitem[{\citenamefont{Dine et~al.}(1995)\citenamefont{Dine, Nelson, and
  Shirman}}]{Dine:1994vc}
\bibinfo{author}{\bibfnamefont{M.}~\bibnamefont{Dine}},
  \bibinfo{author}{\bibfnamefont{A.~E.} \bibnamefont{Nelson}},
  \bibnamefont{and} \bibinfo{author}{\bibfnamefont{Y.}~\bibnamefont{Shirman}},
  \bibinfo{journal}{Phys.Rev.} \textbf{\bibinfo{volume}{D51}},
  \bibinfo{pages}{1362} (\bibinfo{year}{1995}), \eprint{hep-ph/9408384}.

\bibitem[{\citenamefont{Dine et~al.}(1996)\citenamefont{Dine, Nelson, Nir, and
  Shirman}}]{Dine:1995ag}
\bibinfo{author}{\bibfnamefont{M.}~\bibnamefont{Dine}},
  \bibinfo{author}{\bibfnamefont{A.~E.} \bibnamefont{Nelson}},
  \bibinfo{author}{\bibfnamefont{Y.}~\bibnamefont{Nir}}, \bibnamefont{and}
  \bibinfo{author}{\bibfnamefont{Y.}~\bibnamefont{Shirman}},
  \bibinfo{journal}{Phys.Rev.} \textbf{\bibinfo{volume}{D53}},
  \bibinfo{pages}{2658} (\bibinfo{year}{1996}), \eprint{hep-ph/9507378}.

\bibitem[{\citenamefont{Giudice and Rattazzi}(1999)}]{Giudice:1998bp}
\bibinfo{author}{\bibfnamefont{G.}~\bibnamefont{Giudice}} \bibnamefont{and}
  \bibinfo{author}{\bibfnamefont{R.}~\bibnamefont{Rattazzi}},
  \bibinfo{journal}{Phys.Rept.} \textbf{\bibinfo{volume}{322}},
  \bibinfo{pages}{419} (\bibinfo{year}{1999}), \eprint{hep-ph/9801271}.

\bibitem[{\citenamefont{Craig and Green}(2009{\natexlab{a}})}]{Craig:2008vs}
\bibinfo{author}{\bibfnamefont{N.~J.} \bibnamefont{Craig}} \bibnamefont{and}
  \bibinfo{author}{\bibfnamefont{D.~R.} \bibnamefont{Green}},
  \bibinfo{journal}{Phys.Rev.} \textbf{\bibinfo{volume}{D79}},
  \bibinfo{pages}{065030} (\bibinfo{year}{2009}{\natexlab{a}}),
  \eprint{0808.1097}.

\bibitem[{\citenamefont{Shirai et~al.}(2008)\citenamefont{Shirai, Takahashi,
  Yanagida, and Yonekura}}]{Shirai:2008qt}
\bibinfo{author}{\bibfnamefont{S.}~\bibnamefont{Shirai}},
  \bibinfo{author}{\bibfnamefont{F.}~\bibnamefont{Takahashi}},
  \bibinfo{author}{\bibfnamefont{T.}~\bibnamefont{Yanagida}}, \bibnamefont{and}
  \bibinfo{author}{\bibfnamefont{K.}~\bibnamefont{Yonekura}},
  \bibinfo{journal}{Phys.Rev.} \textbf{\bibinfo{volume}{D78}},
  \bibinfo{pages}{075003} (\bibinfo{year}{2008}), \eprint{0808.0848}.

\bibitem[{\citenamefont{Endo et~al.}(2010)\citenamefont{Endo, Shirai, and
  Yonekura}}]{Endo:2009uj}
\bibinfo{author}{\bibfnamefont{M.}~\bibnamefont{Endo}},
  \bibinfo{author}{\bibfnamefont{S.}~\bibnamefont{Shirai}}, \bibnamefont{and}
  \bibinfo{author}{\bibfnamefont{K.}~\bibnamefont{Yonekura}},
  \bibinfo{journal}{JHEP} \textbf{\bibinfo{volume}{1003}}, \bibinfo{pages}{052}
  (\bibinfo{year}{2010}), \eprint{0912.4484}.

\bibitem[{\citenamefont{Craig and Green}(2009{\natexlab{b}})}]{Craig:2009tz}
\bibinfo{author}{\bibfnamefont{N.~J.} \bibnamefont{Craig}} \bibnamefont{and}
  \bibinfo{author}{\bibfnamefont{D.}~\bibnamefont{Green}},
  \bibinfo{journal}{Phys.Rev.} \textbf{\bibinfo{volume}{D80}},
  \bibinfo{pages}{085012} (\bibinfo{year}{2009}{\natexlab{b}}),
  \eprint{0906.2022}.

\bibitem[{\citenamefont{Aad et~al.}(2012)}]{Aad:2012tfa}
\bibinfo{author}{\bibfnamefont{G.}~\bibnamefont{Aad}} \bibnamefont{et~al.}
  (\bibinfo{collaboration}{ATLAS Collaboration}), \bibinfo{journal}{Phys.Lett.}
  \textbf{\bibinfo{volume}{B716}}, \bibinfo{pages}{1} (\bibinfo{year}{2012}),
  \eprint{1207.7214}.

\bibitem[{\citenamefont{Chatrchyan et~al.}(2012)}]{Chatrchyan:2012ufa}
\bibinfo{author}{\bibfnamefont{S.}~\bibnamefont{Chatrchyan}}
  \bibnamefont{et~al.} (\bibinfo{collaboration}{CMS Collaboration}),
  \bibinfo{journal}{Phys.Lett.} \textbf{\bibinfo{volume}{B716}},
  \bibinfo{pages}{30} (\bibinfo{year}{2012}), \eprint{1207.7235}.

\bibitem[{\citenamefont{Carena et~al.}(2012)\citenamefont{Carena, Gori, Shah,
  and Wagner}}]{Carena:2011aa}
\bibinfo{author}{\bibfnamefont{M.}~\bibnamefont{Carena}},
  \bibinfo{author}{\bibfnamefont{S.}~\bibnamefont{Gori}},
  \bibinfo{author}{\bibfnamefont{N.~R.} \bibnamefont{Shah}}, \bibnamefont{and}
  \bibinfo{author}{\bibfnamefont{C.~E.} \bibnamefont{Wagner}},
  \bibinfo{journal}{JHEP} \textbf{\bibinfo{volume}{1203}}, \bibinfo{pages}{014}
  (\bibinfo{year}{2012}), \eprint{1112.3336}.

\bibitem[{\citenamefont{Hall et~al.}(2012)\citenamefont{Hall, Pinner, and
  Ruderman}}]{Hall:2011aa}
\bibinfo{author}{\bibfnamefont{L.~J.} \bibnamefont{Hall}},
  \bibinfo{author}{\bibfnamefont{D.}~\bibnamefont{Pinner}}, \bibnamefont{and}
  \bibinfo{author}{\bibfnamefont{J.~T.} \bibnamefont{Ruderman}},
  \bibinfo{journal}{JHEP} \textbf{\bibinfo{volume}{1204}}, \bibinfo{pages}{131}
  (\bibinfo{year}{2012}), \eprint{1112.2703}.

\bibitem[{\citenamefont{Draper et~al.}(2012)\citenamefont{Draper, Meade, Reece,
  and Shih}}]{Draper:2011aa}
\bibinfo{author}{\bibfnamefont{P.}~\bibnamefont{Draper}},
  \bibinfo{author}{\bibfnamefont{P.}~\bibnamefont{Meade}},
  \bibinfo{author}{\bibfnamefont{M.}~\bibnamefont{Reece}}, \bibnamefont{and}
  \bibinfo{author}{\bibfnamefont{D.}~\bibnamefont{Shih}},
  \bibinfo{journal}{Phys.Rev.} \textbf{\bibinfo{volume}{D85}},
  \bibinfo{pages}{095007} (\bibinfo{year}{2012}), \eprint{1112.3068}.

\bibitem[{\citenamefont{Baer et~al.}(2012{\natexlab{a}})\citenamefont{Baer,
  Barger, and Mustafayev}}]{Baer:2011ab}
\bibinfo{author}{\bibfnamefont{H.}~\bibnamefont{Baer}},
  \bibinfo{author}{\bibfnamefont{V.}~\bibnamefont{Barger}}, \bibnamefont{and}
  \bibinfo{author}{\bibfnamefont{A.}~\bibnamefont{Mustafayev}},
  \bibinfo{journal}{Phys.Rev.} \textbf{\bibinfo{volume}{D85}},
  \bibinfo{pages}{075010} (\bibinfo{year}{2012}{\natexlab{a}}),
  \eprint{1112.3017}.

\bibitem[{\citenamefont{Li et~al.}(2012)\citenamefont{Li, Maxin, Nanopoulos,
  and Walker}}]{Li:2011ab}
\bibinfo{author}{\bibfnamefont{T.}~\bibnamefont{Li}},
  \bibinfo{author}{\bibfnamefont{J.~A.} \bibnamefont{Maxin}},
  \bibinfo{author}{\bibfnamefont{D.~V.} \bibnamefont{Nanopoulos}},
  \bibnamefont{and} \bibinfo{author}{\bibfnamefont{J.~W.}
  \bibnamefont{Walker}}, \bibinfo{journal}{Phys.Lett.}
  \textbf{\bibinfo{volume}{B710}}, \bibinfo{pages}{207} (\bibinfo{year}{2012}),
  \eprint{1112.3024}.

\bibitem[{\citenamefont{Heinemeyer et~al.}(2012)\citenamefont{Heinemeyer, Stal,
  and Weiglein}}]{Heinemeyer:2011aa}
\bibinfo{author}{\bibfnamefont{S.}~\bibnamefont{Heinemeyer}},
  \bibinfo{author}{\bibfnamefont{O.}~\bibnamefont{Stal}}, \bibnamefont{and}
  \bibinfo{author}{\bibfnamefont{G.}~\bibnamefont{Weiglein}},
  \bibinfo{journal}{Phys.Lett.} \textbf{\bibinfo{volume}{B710}},
  \bibinfo{pages}{201} (\bibinfo{year}{2012}), \eprint{1112.3026}.

\bibitem[{\citenamefont{Arbey et~al.}(2012)\citenamefont{Arbey, Battaglia,
  Djouadi, Mahmoudi, and Quevillon}}]{Arbey:2011ab}
\bibinfo{author}{\bibfnamefont{A.}~\bibnamefont{Arbey}},
  \bibinfo{author}{\bibfnamefont{M.}~\bibnamefont{Battaglia}},
  \bibinfo{author}{\bibfnamefont{A.}~\bibnamefont{Djouadi}},
  \bibinfo{author}{\bibfnamefont{F.}~\bibnamefont{Mahmoudi}}, \bibnamefont{and}
  \bibinfo{author}{\bibfnamefont{J.}~\bibnamefont{Quevillon}},
  \bibinfo{journal}{Phys.Lett.} \textbf{\bibinfo{volume}{B708}},
  \bibinfo{pages}{162} (\bibinfo{year}{2012}), \eprint{1112.3028}.

\bibitem[{\citenamefont{Kang et~al.}(2012{\natexlab{a}})\citenamefont{Kang, Li,
  and Li}}]{Kang:2012sy}
\bibinfo{author}{\bibfnamefont{Z.}~\bibnamefont{Kang}},
  \bibinfo{author}{\bibfnamefont{J.}~\bibnamefont{Li}}, \bibnamefont{and}
  \bibinfo{author}{\bibfnamefont{T.}~\bibnamefont{Li}}, \bibinfo{journal}{JHEP}
  \textbf{\bibinfo{volume}{1211}}, \bibinfo{pages}{024}
  (\bibinfo{year}{2012}{\natexlab{a}}), \eprint{1201.5305}.

\bibitem[{\citenamefont{Cao et~al.}(2012)\citenamefont{Cao, Heng, Yang, Zhang,
  and Zhu}}]{Cao:2012fz}
\bibinfo{author}{\bibfnamefont{J.-J.} \bibnamefont{Cao}},
  \bibinfo{author}{\bibfnamefont{Z.-X.} \bibnamefont{Heng}},
  \bibinfo{author}{\bibfnamefont{J.~M.} \bibnamefont{Yang}},
  \bibinfo{author}{\bibfnamefont{Y.-M.} \bibnamefont{Zhang}}, \bibnamefont{and}
  \bibinfo{author}{\bibfnamefont{J.-Y.} \bibnamefont{Zhu}},
  \bibinfo{journal}{JHEP} \textbf{\bibinfo{volume}{1203}}, \bibinfo{pages}{086}
  (\bibinfo{year}{2012}), \eprint{1202.5821}.

\bibitem[{\citenamefont{Ajaib et~al.}(2012)\citenamefont{Ajaib, Gogoladze,
  Nasir, and Shafi}}]{Ajaib:2012vc}
\bibinfo{author}{\bibfnamefont{M.~A.} \bibnamefont{Ajaib}},
  \bibinfo{author}{\bibfnamefont{I.}~\bibnamefont{Gogoladze}},
  \bibinfo{author}{\bibfnamefont{F.}~\bibnamefont{Nasir}}, \bibnamefont{and}
  \bibinfo{author}{\bibfnamefont{Q.}~\bibnamefont{Shafi}},
  \bibinfo{journal}{Phys.Lett.} \textbf{\bibinfo{volume}{B713}},
  \bibinfo{pages}{462} (\bibinfo{year}{2012}), \eprint{1204.2856}.

\bibitem[{\citenamefont{Ke et~al.}(2013)\citenamefont{Ke, Luo, Shan, Wang, and
  Wang}}]{Ke:2012qc}
\bibinfo{author}{\bibfnamefont{J.}~\bibnamefont{Ke}},
  \bibinfo{author}{\bibfnamefont{M.-X.} \bibnamefont{Luo}},
  \bibinfo{author}{\bibfnamefont{L.-Y.} \bibnamefont{Shan}},
  \bibinfo{author}{\bibfnamefont{K.}~\bibnamefont{Wang}}, \bibnamefont{and}
  \bibinfo{author}{\bibfnamefont{L.}~\bibnamefont{Wang}},
  \bibinfo{journal}{Phys.Lett.} \textbf{\bibinfo{volume}{B718}},
  \bibinfo{pages}{1334} (\bibinfo{year}{2013}), \eprint{1207.0990}.

\bibitem[{\citenamefont{Barbieri and Giudice}(1988)}]{Barbieri:1987fn}
\bibinfo{author}{\bibfnamefont{R.}~\bibnamefont{Barbieri}} \bibnamefont{and}
  \bibinfo{author}{\bibfnamefont{G.}~\bibnamefont{Giudice}},
  \bibinfo{journal}{Nucl.Phys.} \textbf{\bibinfo{volume}{B306}},
  \bibinfo{pages}{63} (\bibinfo{year}{1988}).

\bibitem[{\citenamefont{Anderson and Castano}(1995)}]{Anderson:1994dz}
\bibinfo{author}{\bibfnamefont{G.~W.} \bibnamefont{Anderson}} \bibnamefont{and}
  \bibinfo{author}{\bibfnamefont{D.~J.} \bibnamefont{Castano}},
  \bibinfo{journal}{Phys.Lett.} \textbf{\bibinfo{volume}{B347}},
  \bibinfo{pages}{300} (\bibinfo{year}{1995}), \eprint{hep-ph/9409419}.

\bibitem[{\citenamefont{Cohen et~al.}(1996)\citenamefont{Cohen, Kaplan, and
  Nelson}}]{Cohen:1996vb}
\bibinfo{author}{\bibfnamefont{A.~G.} \bibnamefont{Cohen}},
  \bibinfo{author}{\bibfnamefont{D.}~\bibnamefont{Kaplan}}, \bibnamefont{and}
  \bibinfo{author}{\bibfnamefont{A.}~\bibnamefont{Nelson}},
  \bibinfo{journal}{Phys.Lett.} \textbf{\bibinfo{volume}{B388}},
  \bibinfo{pages}{588} (\bibinfo{year}{1996}), \eprint{hep-ph/9607394}.

\bibitem[{\citenamefont{Ciafaloni and Strumia}(1997)}]{Ciafaloni:1996zh}
\bibinfo{author}{\bibfnamefont{P.}~\bibnamefont{Ciafaloni}} \bibnamefont{and}
  \bibinfo{author}{\bibfnamefont{A.}~\bibnamefont{Strumia}},
  \bibinfo{journal}{Nucl.Phys.} \textbf{\bibinfo{volume}{B494}},
  \bibinfo{pages}{41} (\bibinfo{year}{1997}), \eprint{hep-ph/9611204}.

\bibitem[{\citenamefont{Bhattacharyya and
  Romanino}(1997)}]{Bhattacharyya:1996dw}
\bibinfo{author}{\bibfnamefont{G.}~\bibnamefont{Bhattacharyya}}
  \bibnamefont{and} \bibinfo{author}{\bibfnamefont{A.}~\bibnamefont{Romanino}},
  \bibinfo{journal}{Phys.Rev.} \textbf{\bibinfo{volume}{D55}},
  \bibinfo{pages}{7015} (\bibinfo{year}{1997}), \eprint{hep-ph/9611243}.

\bibitem[{\citenamefont{Chankowski et~al.}(1998)\citenamefont{Chankowski,
  Ellis, and Pokorski}}]{Chankowski:1997zh}
\bibinfo{author}{\bibfnamefont{P.~H.} \bibnamefont{Chankowski}},
  \bibinfo{author}{\bibfnamefont{J.~R.} \bibnamefont{Ellis}}, \bibnamefont{and}
  \bibinfo{author}{\bibfnamefont{S.}~\bibnamefont{Pokorski}},
  \bibinfo{journal}{Phys.Lett.} \textbf{\bibinfo{volume}{B423}},
  \bibinfo{pages}{327} (\bibinfo{year}{1998}), \eprint{hep-ph/9712234}.

\bibitem[{\citenamefont{Barbieri and Strumia}(1998)}]{Barbieri:1998uv}
\bibinfo{author}{\bibfnamefont{R.}~\bibnamefont{Barbieri}} \bibnamefont{and}
  \bibinfo{author}{\bibfnamefont{A.}~\bibnamefont{Strumia}},
  \bibinfo{journal}{Phys.Lett.} \textbf{\bibinfo{volume}{B433}},
  \bibinfo{pages}{63} (\bibinfo{year}{1998}), \eprint{hep-ph/9801353}.

\bibitem[{\citenamefont{Kane and King}(1999)}]{Kane:1998im}
\bibinfo{author}{\bibfnamefont{G.~L.} \bibnamefont{Kane}} \bibnamefont{and}
  \bibinfo{author}{\bibfnamefont{S.}~\bibnamefont{King}},
  \bibinfo{journal}{Phys.Lett.} \textbf{\bibinfo{volume}{B451}},
  \bibinfo{pages}{113} (\bibinfo{year}{1999}), \eprint{hep-ph/9810374}.

\bibitem[{\citenamefont{Giusti et~al.}(1999)\citenamefont{Giusti, Romanino, and
  Strumia}}]{Giusti:1998gz}
\bibinfo{author}{\bibfnamefont{L.}~\bibnamefont{Giusti}},
  \bibinfo{author}{\bibfnamefont{A.}~\bibnamefont{Romanino}}, \bibnamefont{and}
  \bibinfo{author}{\bibfnamefont{A.}~\bibnamefont{Strumia}},
  \bibinfo{journal}{Nucl.Phys.} \textbf{\bibinfo{volume}{B550}},
  \bibinfo{pages}{3} (\bibinfo{year}{1999}), \eprint{hep-ph/9811386}.

\bibitem[{\citenamefont{Bastero-Gil et~al.}(2000)\citenamefont{Bastero-Gil,
  Kane, and King}}]{BasteroGil:1999gu}
\bibinfo{author}{\bibfnamefont{M.}~\bibnamefont{Bastero-Gil}},
  \bibinfo{author}{\bibfnamefont{G.~L.} \bibnamefont{Kane}}, \bibnamefont{and}
  \bibinfo{author}{\bibfnamefont{S.}~\bibnamefont{King}},
  \bibinfo{journal}{Phys.Lett.} \textbf{\bibinfo{volume}{B474}},
  \bibinfo{pages}{103} (\bibinfo{year}{2000}), \eprint{hep-ph/9910506}.

\bibitem[{\citenamefont{Feng et~al.}(2000{\natexlab{a}})\citenamefont{Feng,
  Matchev, and Moroi}}]{Feng:1999mn}
\bibinfo{author}{\bibfnamefont{J.~L.} \bibnamefont{Feng}},
  \bibinfo{author}{\bibfnamefont{K.~T.} \bibnamefont{Matchev}},
  \bibnamefont{and} \bibinfo{author}{\bibfnamefont{T.}~\bibnamefont{Moroi}},
  \bibinfo{journal}{Phys.Rev.Lett.} \textbf{\bibinfo{volume}{84}},
  \bibinfo{pages}{2322} (\bibinfo{year}{2000}{\natexlab{a}}),
  \eprint{hep-ph/9908309}.

\bibitem[{\citenamefont{Romanino and Strumia}(2000)}]{Romanino:1999ut}
\bibinfo{author}{\bibfnamefont{A.}~\bibnamefont{Romanino}} \bibnamefont{and}
  \bibinfo{author}{\bibfnamefont{A.}~\bibnamefont{Strumia}},
  \bibinfo{journal}{Phys.Lett.} \textbf{\bibinfo{volume}{B487}},
  \bibinfo{pages}{165} (\bibinfo{year}{2000}), \eprint{hep-ph/9912301}.

\bibitem[{\citenamefont{Feng et~al.}(2000{\natexlab{b}})\citenamefont{Feng,
  Matchev, and Moroi}}]{Feng:1999zg}
\bibinfo{author}{\bibfnamefont{J.~L.} \bibnamefont{Feng}},
  \bibinfo{author}{\bibfnamefont{K.~T.} \bibnamefont{Matchev}},
  \bibnamefont{and} \bibinfo{author}{\bibfnamefont{T.}~\bibnamefont{Moroi}},
  \bibinfo{journal}{Phys.Rev.} \textbf{\bibinfo{volume}{D61}},
  \bibinfo{pages}{075005} (\bibinfo{year}{2000}{\natexlab{b}}),
  \eprint{hep-ph/9909334}.

\bibitem[{\citenamefont{Chacko et~al.}(2005)\citenamefont{Chacko, Nomura, and
  Tucker-Smith}}]{Chacko:2005ra}
\bibinfo{author}{\bibfnamefont{Z.}~\bibnamefont{Chacko}},
  \bibinfo{author}{\bibfnamefont{Y.}~\bibnamefont{Nomura}}, \bibnamefont{and}
  \bibinfo{author}{\bibfnamefont{D.}~\bibnamefont{Tucker-Smith}},
  \bibinfo{journal}{Nucl.Phys.} \textbf{\bibinfo{volume}{B725}},
  \bibinfo{pages}{207} (\bibinfo{year}{2005}), \eprint{hep-ph/0504095}.

\bibitem[{\citenamefont{Choi et~al.}(2006)\citenamefont{Choi, Jeong, Kobayashi,
  and Okumura}}]{Choi:2005hd}
\bibinfo{author}{\bibfnamefont{K.}~\bibnamefont{Choi}},
  \bibinfo{author}{\bibfnamefont{K.~S.} \bibnamefont{Jeong}},
  \bibinfo{author}{\bibfnamefont{T.}~\bibnamefont{Kobayashi}},
  \bibnamefont{and} \bibinfo{author}{\bibfnamefont{K.-i.}
  \bibnamefont{Okumura}}, \bibinfo{journal}{Phys.Lett.}
  \textbf{\bibinfo{volume}{B633}}, \bibinfo{pages}{355} (\bibinfo{year}{2006}),
  \eprint{hep-ph/0508029}.

\bibitem[{\citenamefont{Nomura and Tweedie}(2005)}]{Nomura:2005qg}
\bibinfo{author}{\bibfnamefont{Y.}~\bibnamefont{Nomura}} \bibnamefont{and}
  \bibinfo{author}{\bibfnamefont{B.}~\bibnamefont{Tweedie}},
  \bibinfo{journal}{Phys.Rev.} \textbf{\bibinfo{volume}{D72}},
  \bibinfo{pages}{015006} (\bibinfo{year}{2005}), \eprint{hep-ph/0504246}.

\bibitem[{\citenamefont{Kitano and Nomura}(2005)}]{Kitano:2005wc}
\bibinfo{author}{\bibfnamefont{R.}~\bibnamefont{Kitano}} \bibnamefont{and}
  \bibinfo{author}{\bibfnamefont{Y.}~\bibnamefont{Nomura}},
  \bibinfo{journal}{Phys.Lett.} \textbf{\bibinfo{volume}{B631}},
  \bibinfo{pages}{58} (\bibinfo{year}{2005}), \eprint{hep-ph/0509039}.

\bibitem[{\citenamefont{Nomura et~al.}(2006)\citenamefont{Nomura, Poland, and
  Tweedie}}]{Nomura:2005rj}
\bibinfo{author}{\bibfnamefont{Y.}~\bibnamefont{Nomura}},
  \bibinfo{author}{\bibfnamefont{D.}~\bibnamefont{Poland}}, \bibnamefont{and}
  \bibinfo{author}{\bibfnamefont{B.}~\bibnamefont{Tweedie}},
  \bibinfo{journal}{Nucl.Phys.} \textbf{\bibinfo{volume}{B745}},
  \bibinfo{pages}{29} (\bibinfo{year}{2006}), \eprint{hep-ph/0509243}.

\bibitem[{\citenamefont{Lebedev et~al.}(2005)\citenamefont{Lebedev, Nilles, and
  Ratz}}]{Lebedev:2005ge}
\bibinfo{author}{\bibfnamefont{O.}~\bibnamefont{Lebedev}},
  \bibinfo{author}{\bibfnamefont{H.~P.} \bibnamefont{Nilles}},
  \bibnamefont{and} \bibinfo{author}{\bibfnamefont{M.}~\bibnamefont{Ratz}}, pp.
  \bibinfo{pages}{211--221} (\bibinfo{year}{2005}), \eprint{hep-ph/0511320}.

\bibitem[{\citenamefont{Kitano and Nomura}(2006)}]{Kitano:2006gv}
\bibinfo{author}{\bibfnamefont{R.}~\bibnamefont{Kitano}} \bibnamefont{and}
  \bibinfo{author}{\bibfnamefont{Y.}~\bibnamefont{Nomura}},
  \bibinfo{journal}{Phys.Rev.} \textbf{\bibinfo{volume}{D73}},
  \bibinfo{pages}{095004} (\bibinfo{year}{2006}), \eprint{hep-ph/0602096}.

\bibitem[{\citenamefont{Allanach}(2006)}]{Allanach:2006jc}
\bibinfo{author}{\bibfnamefont{B.}~\bibnamefont{Allanach}},
  \bibinfo{journal}{Phys.Lett.} \textbf{\bibinfo{volume}{B635}},
  \bibinfo{pages}{123} (\bibinfo{year}{2006}), \eprint{hep-ph/0601089}.

\bibitem[{\citenamefont{Giudice and Rattazzi}(2006)}]{Giudice:2006sn}
\bibinfo{author}{\bibfnamefont{G.}~\bibnamefont{Giudice}} \bibnamefont{and}
  \bibinfo{author}{\bibfnamefont{R.}~\bibnamefont{Rattazzi}},
  \bibinfo{journal}{Nucl.Phys.} \textbf{\bibinfo{volume}{B757}},
  \bibinfo{pages}{19} (\bibinfo{year}{2006}), \eprint{hep-ph/0606105}.

\bibitem[{\citenamefont{Perelstein and Spethmann}(2007)}]{Perelstein:2007nx}
\bibinfo{author}{\bibfnamefont{M.}~\bibnamefont{Perelstein}} \bibnamefont{and}
  \bibinfo{author}{\bibfnamefont{C.}~\bibnamefont{Spethmann}},
  \bibinfo{journal}{JHEP} \textbf{\bibinfo{volume}{0704}}, \bibinfo{pages}{070}
  (\bibinfo{year}{2007}), \eprint{hep-ph/0702038}.

\bibitem[{\citenamefont{Allanach et~al.}(2007)\citenamefont{Allanach, Cranmer,
  Lester, and Weber}}]{Allanach:2007qk}
\bibinfo{author}{\bibfnamefont{B.~C.} \bibnamefont{Allanach}},
  \bibinfo{author}{\bibfnamefont{K.}~\bibnamefont{Cranmer}},
  \bibinfo{author}{\bibfnamefont{C.~G.} \bibnamefont{Lester}},
  \bibnamefont{and} \bibinfo{author}{\bibfnamefont{A.~M.} \bibnamefont{Weber}},
  \bibinfo{journal}{JHEP} \textbf{\bibinfo{volume}{0708}}, \bibinfo{pages}{023}
  (\bibinfo{year}{2007}), \eprint{0705.0487}.

\bibitem[{\citenamefont{Cabrera et~al.}(2009)\citenamefont{Cabrera, Casas, and
  Ruiz~de Austri}}]{Cabrera:2008tj}
\bibinfo{author}{\bibfnamefont{M.}~\bibnamefont{Cabrera}},
  \bibinfo{author}{\bibfnamefont{J.}~\bibnamefont{Casas}}, \bibnamefont{and}
  \bibinfo{author}{\bibfnamefont{R.}~\bibnamefont{Ruiz~de Austri}},
  \bibinfo{journal}{JHEP} \textbf{\bibinfo{volume}{0903}}, \bibinfo{pages}{075}
  (\bibinfo{year}{2009}), \eprint{0812.0536}.

\bibitem[{\citenamefont{Cassel et~al.}(2010)\citenamefont{Cassel, Ghilencea,
  and Ross}}]{Cassel:2009ps}
\bibinfo{author}{\bibfnamefont{S.}~\bibnamefont{Cassel}},
  \bibinfo{author}{\bibfnamefont{D.}~\bibnamefont{Ghilencea}},
  \bibnamefont{and} \bibinfo{author}{\bibfnamefont{G.}~\bibnamefont{Ross}},
  \bibinfo{journal}{Nucl.Phys.} \textbf{\bibinfo{volume}{B825}},
  \bibinfo{pages}{203} (\bibinfo{year}{2010}), \eprint{0903.1115}.

\bibitem[{\citenamefont{Barbieri and Pappadopulo}(2009)}]{Barbieri:2009ev}
\bibinfo{author}{\bibfnamefont{R.}~\bibnamefont{Barbieri}} \bibnamefont{and}
  \bibinfo{author}{\bibfnamefont{D.}~\bibnamefont{Pappadopulo}},
  \bibinfo{journal}{JHEP} \textbf{\bibinfo{volume}{0910}}, \bibinfo{pages}{061}
  (\bibinfo{year}{2009}), \eprint{0906.4546}.

\bibitem[{\citenamefont{Horton and Ross}(2010)}]{Horton:2009ed}
\bibinfo{author}{\bibfnamefont{D.}~\bibnamefont{Horton}} \bibnamefont{and}
  \bibinfo{author}{\bibfnamefont{G.}~\bibnamefont{Ross}},
  \bibinfo{journal}{Nucl.Phys.} \textbf{\bibinfo{volume}{B830}},
  \bibinfo{pages}{221} (\bibinfo{year}{2010}), \eprint{0908.0857}.

\bibitem[{\citenamefont{Kobayashi et~al.}(2010)\citenamefont{Kobayashi, Nakai,
  and Takahashi}}]{Kobayashi:2009rn}
\bibinfo{author}{\bibfnamefont{T.}~\bibnamefont{Kobayashi}},
  \bibinfo{author}{\bibfnamefont{Y.}~\bibnamefont{Nakai}}, \bibnamefont{and}
  \bibinfo{author}{\bibfnamefont{R.}~\bibnamefont{Takahashi}},
  \bibinfo{journal}{JHEP} \textbf{\bibinfo{volume}{1001}}, \bibinfo{pages}{003}
  (\bibinfo{year}{2010}), \eprint{0910.3477}.

\bibitem[{\citenamefont{Lodone}(2010)}]{Lodone:2010kt}
\bibinfo{author}{\bibfnamefont{P.}~\bibnamefont{Lodone}},
  \bibinfo{journal}{JHEP} \textbf{\bibinfo{volume}{1005}}, \bibinfo{pages}{068}
  (\bibinfo{year}{2010}), \eprint{1004.1271}.

\bibitem[{\citenamefont{Asano et~al.}(2010)\citenamefont{Asano, Kim, Kitano,
  and Shimizu}}]{Asano:2010ut}
\bibinfo{author}{\bibfnamefont{M.}~\bibnamefont{Asano}},
  \bibinfo{author}{\bibfnamefont{H.~D.} \bibnamefont{Kim}},
  \bibinfo{author}{\bibfnamefont{R.}~\bibnamefont{Kitano}}, \bibnamefont{and}
  \bibinfo{author}{\bibfnamefont{Y.}~\bibnamefont{Shimizu}},
  \bibinfo{journal}{JHEP} \textbf{\bibinfo{volume}{1012}}, \bibinfo{pages}{019}
  (\bibinfo{year}{2010}), \eprint{1010.0692}.

\bibitem[{\citenamefont{Strumia}(2011)}]{Strumia:2011dv}
\bibinfo{author}{\bibfnamefont{A.}~\bibnamefont{Strumia}},
  \bibinfo{journal}{JHEP} \textbf{\bibinfo{volume}{1104}}, \bibinfo{pages}{073}
  (\bibinfo{year}{2011}), \eprint{1101.2195}.

\bibitem[{\citenamefont{Cassel et~al.}(2011)\citenamefont{Cassel, Ghilencea,
  Kraml, Lessa, and Ross}}]{Cassel:2011tg}
\bibinfo{author}{\bibfnamefont{S.}~\bibnamefont{Cassel}},
  \bibinfo{author}{\bibfnamefont{D.}~\bibnamefont{Ghilencea}},
  \bibinfo{author}{\bibfnamefont{S.}~\bibnamefont{Kraml}},
  \bibinfo{author}{\bibfnamefont{A.}~\bibnamefont{Lessa}}, \bibnamefont{and}
  \bibinfo{author}{\bibfnamefont{G.}~\bibnamefont{Ross}},
  \bibinfo{journal}{JHEP} \textbf{\bibinfo{volume}{1105}}, \bibinfo{pages}{120}
  (\bibinfo{year}{2011}), \eprint{1101.4664}.

\bibitem[{\citenamefont{Sakurai and Takayama}(2011)}]{Sakurai:2011pt}
\bibinfo{author}{\bibfnamefont{K.}~\bibnamefont{Sakurai}} \bibnamefont{and}
  \bibinfo{author}{\bibfnamefont{K.}~\bibnamefont{Takayama}},
  \bibinfo{journal}{JHEP} \textbf{\bibinfo{volume}{1112}}, \bibinfo{pages}{063}
  (\bibinfo{year}{2011}), \eprint{1106.3794}.

\bibitem[{\citenamefont{Papucci et~al.}(2012)\citenamefont{Papucci, Ruderman,
  and Weiler}}]{Papucci:2011wy}
\bibinfo{author}{\bibfnamefont{M.}~\bibnamefont{Papucci}},
  \bibinfo{author}{\bibfnamefont{J.~T.} \bibnamefont{Ruderman}},
  \bibnamefont{and} \bibinfo{author}{\bibfnamefont{A.}~\bibnamefont{Weiler}},
  \bibinfo{journal}{JHEP} \textbf{\bibinfo{volume}{1209}}, \bibinfo{pages}{035}
  (\bibinfo{year}{2012}), \eprint{1110.6926}.

\bibitem[{\citenamefont{Larsen et~al.}(2012)\citenamefont{Larsen, Nomura, and
  Roberts}}]{Larsen:2012rq}
\bibinfo{author}{\bibfnamefont{G.}~\bibnamefont{Larsen}},
  \bibinfo{author}{\bibfnamefont{Y.}~\bibnamefont{Nomura}}, \bibnamefont{and}
  \bibinfo{author}{\bibfnamefont{H.~L.} \bibnamefont{Roberts}},
  \bibinfo{journal}{JHEP} \textbf{\bibinfo{volume}{1206}}, \bibinfo{pages}{032}
  (\bibinfo{year}{2012}), \eprint{1202.6339}.

\bibitem[{\citenamefont{Baer et~al.}(2012{\natexlab{b}})\citenamefont{Baer,
  Barger, Huang, and Tata}}]{Baer:2012uy}
\bibinfo{author}{\bibfnamefont{H.}~\bibnamefont{Baer}},
  \bibinfo{author}{\bibfnamefont{V.}~\bibnamefont{Barger}},
  \bibinfo{author}{\bibfnamefont{P.}~\bibnamefont{Huang}}, \bibnamefont{and}
  \bibinfo{author}{\bibfnamefont{X.}~\bibnamefont{Tata}},
  \bibinfo{journal}{JHEP} \textbf{\bibinfo{volume}{1205}}, \bibinfo{pages}{109}
  (\bibinfo{year}{2012}{\natexlab{b}}), \eprint{1203.5539}.

\bibitem[{\citenamefont{Espinosa et~al.}(2012)\citenamefont{Espinosa, Grojean,
  Sanz, and Trott}}]{Espinosa:2012in}
\bibinfo{author}{\bibfnamefont{J.~R.} \bibnamefont{Espinosa}},
  \bibinfo{author}{\bibfnamefont{C.}~\bibnamefont{Grojean}},
  \bibinfo{author}{\bibfnamefont{V.}~\bibnamefont{Sanz}}, \bibnamefont{and}
  \bibinfo{author}{\bibfnamefont{M.}~\bibnamefont{Trott}},
  \bibinfo{journal}{JHEP} \textbf{\bibinfo{volume}{1212}}, \bibinfo{pages}{077}
  (\bibinfo{year}{2012}), \eprint{1207.7355}.

\bibitem[{\citenamefont{Boehm et~al.}(2013)\citenamefont{Boehm, Dev, Mazumdar,
  and Pukartas}}]{Boehm:2013qva}
\bibinfo{author}{\bibfnamefont{C.}~\bibnamefont{Boehm}},
  \bibinfo{author}{\bibfnamefont{P.~S.~B.} \bibnamefont{Dev}},
  \bibinfo{author}{\bibfnamefont{A.}~\bibnamefont{Mazumdar}}, \bibnamefont{and}
  \bibinfo{author}{\bibfnamefont{E.}~\bibnamefont{Pukartas}},
  \bibinfo{journal}{JHEP} \textbf{\bibinfo{volume}{1306}}, \bibinfo{pages}{113}
  (\bibinfo{year}{2013}), \eprint{1303.5386}.

\bibitem[{\citenamefont{Zheng}(2013{\natexlab{a}})}]{Zheng:2013afa}
\bibinfo{author}{\bibfnamefont{S.}~\bibnamefont{Zheng}}
  (\bibinfo{year}{2013}{\natexlab{a}}), \eprint{1312.0181}.

\bibitem[{\citenamefont{Zheng}(2013{\natexlab{b}})}]{Zheng:2013kwa}
\bibinfo{author}{\bibfnamefont{S.}~\bibnamefont{Zheng}}
  (\bibinfo{year}{2013}{\natexlab{b}}), \eprint{1312.4105}.

\bibitem[{\citenamefont{Akerib et~al.}(2013)}]{Akerib:2013tjd}
\bibinfo{author}{\bibfnamefont{D.}~\bibnamefont{Akerib}} \bibnamefont{et~al.}
  (\bibinfo{collaboration}{LUX Collaboration}) (\bibinfo{year}{2013}),
  \eprint{1310.8214}.

\bibitem[{\citenamefont{Cao et~al.}(2013)\citenamefont{Cao, Han, Wu, Wu, and
  Yang}}]{Cao:2013mqa}
\bibinfo{author}{\bibfnamefont{J.}~\bibnamefont{Cao}},
  \bibinfo{author}{\bibfnamefont{C.}~\bibnamefont{Han}},
  \bibinfo{author}{\bibfnamefont{L.}~\bibnamefont{Wu}},
  \bibinfo{author}{\bibfnamefont{P.}~\bibnamefont{Wu}}, \bibnamefont{and}
  \bibinfo{author}{\bibfnamefont{J.~M.} \bibnamefont{Yang}}
  (\bibinfo{year}{2013}), \eprint{1311.0678}.

\bibitem[{\citenamefont{Ellis}(2013)}]{Ellis:2013oxa}
\bibinfo{author}{\bibfnamefont{J.}~\bibnamefont{Ellis}} (\bibinfo{year}{2013}),
  \eprint{1312.5426}.

\bibitem[{\citenamefont{Buchmueller
  et~al.}(2013{\natexlab{a}})\citenamefont{Buchmueller, Cavanaugh, De~Roeck,
  Dolan, Ellis et~al.}}]{Buchmueller:2013rsa}
\bibinfo{author}{\bibfnamefont{O.}~\bibnamefont{Buchmueller}},
  \bibinfo{author}{\bibfnamefont{R.}~\bibnamefont{Cavanaugh}},
  \bibinfo{author}{\bibfnamefont{A.}~\bibnamefont{De~Roeck}},
  \bibinfo{author}{\bibfnamefont{M.}~\bibnamefont{Dolan}},
  \bibinfo{author}{\bibfnamefont{J.}~\bibnamefont{Ellis}}, \bibnamefont{et~al.}
  (\bibinfo{year}{2013}{\natexlab{a}}), \eprint{1312.5250}.

\bibitem[{\citenamefont{Martin}(2014)}]{Martin:2013aha}
\bibinfo{author}{\bibfnamefont{S.~P.} \bibnamefont{Martin}},
  \bibinfo{journal}{Phys.Rev.} \textbf{\bibinfo{volume}{D89}},
  \bibinfo{pages}{035011} (\bibinfo{year}{2014}), \eprint{1312.0582}.

\bibitem[{\citenamefont{Guo et~al.}(2013)\citenamefont{Guo, Kang, Li, Li, and
  Liu}}]{Guo:2013asa}
\bibinfo{author}{\bibfnamefont{J.}~\bibnamefont{Guo}},
  \bibinfo{author}{\bibfnamefont{Z.}~\bibnamefont{Kang}},
  \bibinfo{author}{\bibfnamefont{J.}~\bibnamefont{Li}},
  \bibinfo{author}{\bibfnamefont{T.}~\bibnamefont{Li}}, \bibnamefont{and}
  \bibinfo{author}{\bibfnamefont{Y.}~\bibnamefont{Liu}} (\bibinfo{year}{2013}),
  \eprint{1312.2821}.

\bibitem[{\citenamefont{Luty and Sundrum}(2002)}]{Luty:2001jh}
\bibinfo{author}{\bibfnamefont{M.~A.} \bibnamefont{Luty}} \bibnamefont{and}
  \bibinfo{author}{\bibfnamefont{R.}~\bibnamefont{Sundrum}},
  \bibinfo{journal}{Phys. Rev.} \textbf{\bibinfo{volume}{D65}},
  \bibinfo{pages}{066004} (\bibinfo{year}{2002}), \eprint{hep-th/0105137}.

\bibitem[{\citenamefont{Luty and Sundrum}(2003)}]{Luty:2001zv}
\bibinfo{author}{\bibfnamefont{M.}~\bibnamefont{Luty}} \bibnamefont{and}
  \bibinfo{author}{\bibfnamefont{R.}~\bibnamefont{Sundrum}},
  \bibinfo{journal}{Phys. Rev.} \textbf{\bibinfo{volume}{D67}},
  \bibinfo{pages}{045007} (\bibinfo{year}{2003}), \eprint{hep-th/0111231}.

\bibitem[{\citenamefont{Dine et~al.}(2004)\citenamefont{Dine, Fox, Gorbatov,
  Shadmi, Shirman et~al.}}]{Dine:2004dv}
\bibinfo{author}{\bibfnamefont{M.}~\bibnamefont{Dine}},
  \bibinfo{author}{\bibfnamefont{P.}~\bibnamefont{Fox}},
  \bibinfo{author}{\bibfnamefont{E.}~\bibnamefont{Gorbatov}},
  \bibinfo{author}{\bibfnamefont{Y.}~\bibnamefont{Shadmi}},
  \bibinfo{author}{\bibfnamefont{Y.}~\bibnamefont{Shirman}},
  \bibnamefont{et~al.}, \bibinfo{journal}{Phys.Rev.}
  \textbf{\bibinfo{volume}{D70}}, \bibinfo{pages}{045023}
  (\bibinfo{year}{2004}), \eprint{hep-ph/0405159}.

\bibitem[{\citenamefont{Cohen et~al.}(2007)\citenamefont{Cohen, Roy, and
  Schmaltz}}]{Cohen:2006qc}
\bibinfo{author}{\bibfnamefont{A.~G.} \bibnamefont{Cohen}},
  \bibinfo{author}{\bibfnamefont{T.~S.} \bibnamefont{Roy}}, \bibnamefont{and}
  \bibinfo{author}{\bibfnamefont{M.}~\bibnamefont{Schmaltz}},
  \bibinfo{journal}{JHEP} \textbf{\bibinfo{volume}{02}}, \bibinfo{pages}{027}
  (\bibinfo{year}{2007}), \eprint{hep-ph/0612100}.

\bibitem[{\citenamefont{Schmaltz and Sundrum}(2006)}]{Schmaltz:2006qs}
\bibinfo{author}{\bibfnamefont{M.}~\bibnamefont{Schmaltz}} \bibnamefont{and}
  \bibinfo{author}{\bibfnamefont{R.}~\bibnamefont{Sundrum}},
  \bibinfo{journal}{JHEP} \textbf{\bibinfo{volume}{0611}}, \bibinfo{pages}{011}
  (\bibinfo{year}{2006}), \eprint{hep-th/0608051}.

\bibitem[{\citenamefont{Murayama et~al.}(2008)\citenamefont{Murayama, Nomura,
  and Poland}}]{Murayama:2007ge}
\bibinfo{author}{\bibfnamefont{H.}~\bibnamefont{Murayama}},
  \bibinfo{author}{\bibfnamefont{Y.}~\bibnamefont{Nomura}}, \bibnamefont{and}
  \bibinfo{author}{\bibfnamefont{D.}~\bibnamefont{Poland}},
  \bibinfo{journal}{Phys. Rev.} \textbf{\bibinfo{volume}{D77}},
  \bibinfo{pages}{015005} (\bibinfo{year}{2008}), \eprint{0709.0775}.

\bibitem[{\citenamefont{Roy and Schmaltz}(2008)}]{Roy:2007nz}
\bibinfo{author}{\bibfnamefont{T.~S.} \bibnamefont{Roy}} \bibnamefont{and}
  \bibinfo{author}{\bibfnamefont{M.}~\bibnamefont{Schmaltz}},
  \bibinfo{journal}{Phys. Rev.} \textbf{\bibinfo{volume}{D77}},
  \bibinfo{pages}{095008} (\bibinfo{year}{2008}), \eprint{0708.3593}.

\bibitem[{\citenamefont{Perez et~al.}(2009)\citenamefont{Perez, Roy, and
  Schmaltz}}]{Perez:2008ng}
\bibinfo{author}{\bibfnamefont{G.}~\bibnamefont{Perez}},
  \bibinfo{author}{\bibfnamefont{T.~S.} \bibnamefont{Roy}}, \bibnamefont{and}
  \bibinfo{author}{\bibfnamefont{M.}~\bibnamefont{Schmaltz}},
  \bibinfo{journal}{Phys.Rev.} \textbf{\bibinfo{volume}{D79}},
  \bibinfo{pages}{095016} (\bibinfo{year}{2009}), \eprint{0811.3206}.

\bibitem[{\citenamefont{Komargodski and Seiberg}(2009)}]{Komargodski:2008ax}
\bibinfo{author}{\bibfnamefont{Z.}~\bibnamefont{Komargodski}} \bibnamefont{and}
  \bibinfo{author}{\bibfnamefont{N.}~\bibnamefont{Seiberg}},
  \bibinfo{journal}{JHEP} \textbf{\bibinfo{volume}{03}}, \bibinfo{pages}{072}
  (\bibinfo{year}{2009}), \eprint{0812.3900}.

\bibitem[{\citenamefont{Cho}(2008)}]{Cho:2008fr}
\bibinfo{author}{\bibfnamefont{H.~Y.} \bibnamefont{Cho}},
  \bibinfo{journal}{JHEP} \textbf{\bibinfo{volume}{07}}, \bibinfo{pages}{069}
  (\bibinfo{year}{2008}), \eprint{0802.1145}.

\bibitem[{\citenamefont{Asano et~al.}(2009)\citenamefont{Asano, Hisano, Okada,
  and Sugiyama}}]{Asano:2008qc}
\bibinfo{author}{\bibfnamefont{M.}~\bibnamefont{Asano}},
  \bibinfo{author}{\bibfnamefont{J.}~\bibnamefont{Hisano}},
  \bibinfo{author}{\bibfnamefont{T.}~\bibnamefont{Okada}}, \bibnamefont{and}
  \bibinfo{author}{\bibfnamefont{S.}~\bibnamefont{Sugiyama}},
  \bibinfo{journal}{Phys.Lett.} \textbf{\bibinfo{volume}{B673}},
  \bibinfo{pages}{146} (\bibinfo{year}{2009}), \eprint{0810.4606}.

\bibitem[{\citenamefont{Craig and Green}(2009{\natexlab{c}})}]{Craig:2009rk}
\bibinfo{author}{\bibfnamefont{N.~J.} \bibnamefont{Craig}} \bibnamefont{and}
  \bibinfo{author}{\bibfnamefont{D.}~\bibnamefont{Green}},
  \bibinfo{journal}{JHEP} \textbf{\bibinfo{volume}{0909}}, \bibinfo{pages}{113}
  (\bibinfo{year}{2009}{\natexlab{c}}), \eprint{0905.4088}.

\bibitem[{\citenamefont{Evans et~al.}(2012)\citenamefont{Evans, Ibe, and
  Yanagida}}]{Evans:2012uf}
\bibinfo{author}{\bibfnamefont{J.~L.} \bibnamefont{Evans}},
  \bibinfo{author}{\bibfnamefont{M.}~\bibnamefont{Ibe}}, \bibnamefont{and}
  \bibinfo{author}{\bibfnamefont{T.~T.} \bibnamefont{Yanagida}},
  \bibinfo{journal}{Phys.Rev.} \textbf{\bibinfo{volume}{D86}},
  \bibinfo{pages}{015017} (\bibinfo{year}{2012}), \eprint{1204.6085}.

\bibitem[{\citenamefont{Craig et~al.}(2013{\natexlab{a}})\citenamefont{Craig,
  Knapen, and Shih}}]{Craig:2013wga}
\bibinfo{author}{\bibfnamefont{N.}~\bibnamefont{Craig}},
  \bibinfo{author}{\bibfnamefont{S.}~\bibnamefont{Knapen}}, \bibnamefont{and}
  \bibinfo{author}{\bibfnamefont{D.}~\bibnamefont{Shih}},
  \bibinfo{journal}{JHEP} \textbf{\bibinfo{volume}{1308}}, \bibinfo{pages}{118}
  (\bibinfo{year}{2013}{\natexlab{a}}), \eprint{1302.2642}.

\bibitem[{\citenamefont{Knapen and Shih}(2013)}]{Knapen:2013zla}
\bibinfo{author}{\bibfnamefont{S.}~\bibnamefont{Knapen}} \bibnamefont{and}
  \bibinfo{author}{\bibfnamefont{D.}~\bibnamefont{Shih}}
  (\bibinfo{year}{2013}), \eprint{1311.7107}.

\bibitem[{\citenamefont{Ding et~al.}(2013)\citenamefont{Ding, Li, Staub, and
  Zhu}}]{Ding:2013pya}
\bibinfo{author}{\bibfnamefont{R.}~\bibnamefont{Ding}},
  \bibinfo{author}{\bibfnamefont{T.}~\bibnamefont{Li}},
  \bibinfo{author}{\bibfnamefont{F.}~\bibnamefont{Staub}}, \bibnamefont{and}
  \bibinfo{author}{\bibfnamefont{B.}~\bibnamefont{Zhu}} (\bibinfo{year}{2013}),
  \eprint{1312.5407}.

\bibitem[{\citenamefont{Dobrev and Petkova}(1985)}]{Dobrev:1985qv}
\bibinfo{author}{\bibfnamefont{V.}~\bibnamefont{Dobrev}} \bibnamefont{and}
  \bibinfo{author}{\bibfnamefont{V.}~\bibnamefont{Petkova}},
  \bibinfo{journal}{Phys.Lett.} \textbf{\bibinfo{volume}{B162}},
  \bibinfo{pages}{127} (\bibinfo{year}{1985}).

\bibitem[{\citenamefont{Evans and Shih}(2013)}]{Evans2013}
\bibinfo{author}{\bibfnamefont{J.~A.} \bibnamefont{Evans}} \bibnamefont{and}
  \bibinfo{author}{\bibfnamefont{D.}~\bibnamefont{Shih}},
  \bibinfo{journal}{JHEP} \textbf{\bibinfo{volume}{1308}}, \bibinfo{pages}{093}
  (\bibinfo{year}{2013}), \eprint{1303.0228}.

\bibitem[{\citenamefont{Staub et~al.}(2012)\citenamefont{Staub, Ohl, Porod, and
  Speckner}}]{oai:arXiv.org:1109.5147}
\bibinfo{author}{\bibfnamefont{F.}~\bibnamefont{Staub}},
  \bibinfo{author}{\bibfnamefont{T.}~\bibnamefont{Ohl}},
  \bibinfo{author}{\bibfnamefont{W.}~\bibnamefont{Porod}}, \bibnamefont{and}
  \bibinfo{author}{\bibfnamefont{C.}~\bibnamefont{Speckner}},
  \bibinfo{journal}{Comput.Phys.Commun.} \textbf{\bibinfo{volume}{183}},
  \bibinfo{pages}{2165} (\bibinfo{year}{2012}), \eprint{1109.5147}.

\bibitem[{\citenamefont{Staub}(2010)}]{oai:arXiv.org:0909.2863}
\bibinfo{author}{\bibfnamefont{F.}~\bibnamefont{Staub}},
  \bibinfo{journal}{Comput.Phys.Commun.} \textbf{\bibinfo{volume}{181}},
  \bibinfo{pages}{1077} (\bibinfo{year}{2010}), \eprint{0909.2863}.

\bibitem[{\citenamefont{Staub}(2011)}]{oai:arXiv.org:1002.0840}
\bibinfo{author}{\bibfnamefont{F.}~\bibnamefont{Staub}},
  \bibinfo{journal}{Comput.Phys.Commun.} \textbf{\bibinfo{volume}{182}},
  \bibinfo{pages}{808} (\bibinfo{year}{2011}), \eprint{1002.0840}.

\bibitem[{\citenamefont{Staub}(2013)}]{staub}
\bibinfo{author}{\bibfnamefont{F.}~\bibnamefont{Staub}},
  \bibinfo{journal}{Computer Physics Communications}
  \textbf{\bibinfo{volume}{184}}, \bibinfo{pages}{pp. 1792}
  (\bibinfo{year}{2013}), \eprint{1207.0906}.

\bibitem[{\citenamefont{Porod}(2003)}]{oai:arXiv.org:hep-ph/0301101}
\bibinfo{author}{\bibfnamefont{W.}~\bibnamefont{Porod}},
  \bibinfo{journal}{Comput.Phys.Commun.} \textbf{\bibinfo{volume}{153}},
  \bibinfo{pages}{275} (\bibinfo{year}{2003}), \eprint{hep-ph/0301101}.

\bibitem[{\citenamefont{Porod and Staub}(2012)}]{oai:arXiv.org:1104.1573}
\bibinfo{author}{\bibfnamefont{W.}~\bibnamefont{Porod}} \bibnamefont{and}
  \bibinfo{author}{\bibfnamefont{F.}~\bibnamefont{Staub}},
  \bibinfo{journal}{Comput.Phys.Commun.} \textbf{\bibinfo{volume}{183}},
  \bibinfo{pages}{2458} (\bibinfo{year}{2012}), \eprint{1104.1573}.

\bibitem[{\citenamefont{Belanger et~al.}(2009)\citenamefont{Belanger, Boudjema,
  Pukhov, and Semenov}}]{oai:arXiv.org:0803.2360}
\bibinfo{author}{\bibfnamefont{G.}~\bibnamefont{Belanger}},
  \bibinfo{author}{\bibfnamefont{F.}~\bibnamefont{Boudjema}},
  \bibinfo{author}{\bibfnamefont{A.}~\bibnamefont{Pukhov}}, \bibnamefont{and}
  \bibinfo{author}{\bibfnamefont{A.}~\bibnamefont{Semenov}},
  \bibinfo{journal}{Comput.Phys.Commun.} \textbf{\bibinfo{volume}{180}},
  \bibinfo{pages}{747} (\bibinfo{year}{2009}), \eprint{0803.2360}.

\bibitem[{\citenamefont{Feng et~al.}(2013)\citenamefont{Feng, Kant, Profumo,
  and Sanford}}]{Feng:2013tvd}
\bibinfo{author}{\bibfnamefont{J.~L.} \bibnamefont{Feng}},
  \bibinfo{author}{\bibfnamefont{P.}~\bibnamefont{Kant}},
  \bibinfo{author}{\bibfnamefont{S.}~\bibnamefont{Profumo}}, \bibnamefont{and}
  \bibinfo{author}{\bibfnamefont{D.}~\bibnamefont{Sanford}},
  \bibinfo{journal}{Phys.Rev.Lett.} \textbf{\bibinfo{volume}{111}},
  \bibinfo{pages}{131802} (\bibinfo{year}{2013}), \eprint{1306.2318}.

\bibitem[{\citenamefont{Buchmueller
  et~al.}(2013{\natexlab{b}})\citenamefont{Buchmueller, Dolan, Ellis, Hahn,
  Heinemeyer et~al.}}]{Buchmueller:2013psa}
\bibinfo{author}{\bibfnamefont{O.}~\bibnamefont{Buchmueller}},
  \bibinfo{author}{\bibfnamefont{M.}~\bibnamefont{Dolan}},
  \bibinfo{author}{\bibfnamefont{J.}~\bibnamefont{Ellis}},
  \bibinfo{author}{\bibfnamefont{T.}~\bibnamefont{Hahn}},
  \bibinfo{author}{\bibfnamefont{S.}~\bibnamefont{Heinemeyer}},
  \bibnamefont{et~al.} (\bibinfo{year}{2013}{\natexlab{b}}),
  \eprint{1312.5233}.

\bibitem[{\citenamefont{Bennett et~al.}(2013)}]{Bennett:2012zja}
\bibinfo{author}{\bibfnamefont{C.}~\bibnamefont{Bennett}} \bibnamefont{et~al.}
  (\bibinfo{collaboration}{WMAP}), \bibinfo{journal}{Astrophys.J.Suppl.}
  \textbf{\bibinfo{volume}{208}}, \bibinfo{pages}{20} (\bibinfo{year}{2013}),
  \eprint{1212.5225}.

\bibitem[{\citenamefont{Arkani-Hamed et~al.}(2006)\citenamefont{Arkani-Hamed,
  Delgado, and Giudice}}]{ArkaniHamed:2006mb}
\bibinfo{author}{\bibfnamefont{N.}~\bibnamefont{Arkani-Hamed}},
  \bibinfo{author}{\bibfnamefont{A.}~\bibnamefont{Delgado}}, \bibnamefont{and}
  \bibinfo{author}{\bibfnamefont{G.}~\bibnamefont{Giudice}},
  \bibinfo{journal}{Nucl.Phys.} \textbf{\bibinfo{volume}{B741}},
  \bibinfo{pages}{108} (\bibinfo{year}{2006}), \eprint{hep-ph/0601041}.

\bibitem[{\citenamefont{Ellis et~al.}(1998)\citenamefont{Ellis, Falk, and
  Olive}}]{Ellis:1998kh}
\bibinfo{author}{\bibfnamefont{J.~R.} \bibnamefont{Ellis}},
  \bibinfo{author}{\bibfnamefont{T.}~\bibnamefont{Falk}}, \bibnamefont{and}
  \bibinfo{author}{\bibfnamefont{K.~A.} \bibnamefont{Olive}},
  \bibinfo{journal}{Phys.Lett.} \textbf{\bibinfo{volume}{B444}},
  \bibinfo{pages}{367} (\bibinfo{year}{1998}), \eprint{hep-ph/9810360}.

\bibitem[{\citenamefont{Kang et~al.}(2012{\natexlab{b}})\citenamefont{Kang, Li,
  Liu, Tong, and Yang}}]{Kang:2012ra}
\bibinfo{author}{\bibfnamefont{Z.}~\bibnamefont{Kang}},
  \bibinfo{author}{\bibfnamefont{T.}~\bibnamefont{Li}},
  \bibinfo{author}{\bibfnamefont{T.}~\bibnamefont{Liu}},
  \bibinfo{author}{\bibfnamefont{C.}~\bibnamefont{Tong}}, \bibnamefont{and}
  \bibinfo{author}{\bibfnamefont{J.~M.} \bibnamefont{Yang}},
  \bibinfo{journal}{Phys.Rev.} \textbf{\bibinfo{volume}{D86}},
  \bibinfo{pages}{095020} (\bibinfo{year}{2012}{\natexlab{b}}),
  \eprint{1203.2336}.

\bibitem[{\citenamefont{Craig et~al.}(2013{\natexlab{b}})\citenamefont{Craig,
  Knapen, Shih, and Zhao}}]{Craig:2012xp}
\bibinfo{author}{\bibfnamefont{N.}~\bibnamefont{Craig}},
  \bibinfo{author}{\bibfnamefont{S.}~\bibnamefont{Knapen}},
  \bibinfo{author}{\bibfnamefont{D.}~\bibnamefont{Shih}}, \bibnamefont{and}
  \bibinfo{author}{\bibfnamefont{Y.}~\bibnamefont{Zhao}},
  \bibinfo{journal}{JHEP} \textbf{\bibinfo{volume}{1303}}, \bibinfo{pages}{154}
  (\bibinfo{year}{2013}{\natexlab{b}}), \eprint{1206.4086}.

\bibitem[{\citenamefont{Albaid and Babu}(2013)}]{Albaid:2012qk}
\bibinfo{author}{\bibfnamefont{A.}~\bibnamefont{Albaid}} \bibnamefont{and}
  \bibinfo{author}{\bibfnamefont{K.}~\bibnamefont{Babu}},
  \bibinfo{journal}{Phys.Rev.} \textbf{\bibinfo{volume}{D88}},
  \bibinfo{pages}{055007} (\bibinfo{year}{2013}), \eprint{1207.1014}.

\bibitem[{\citenamefont{Zheng}(2014)}]{Zheng:2013lga}
\bibinfo{author}{\bibfnamefont{S.}~\bibnamefont{Zheng}},
  \bibinfo{journal}{Eur.Phys.J.} \textbf{\bibinfo{volume}{C74}},
  \bibinfo{pages}{2724} (\bibinfo{year}{2014}), \eprint{1308.5377}.

\bibitem[{\citenamefont{Byakti and Ray}(2013)}]{Byakti:2013ti}
\bibinfo{author}{\bibfnamefont{P.}~\bibnamefont{Byakti}} \bibnamefont{and}
  \bibinfo{author}{\bibfnamefont{T.~S.} \bibnamefont{Ray}},
  \bibinfo{journal}{JHEP} \textbf{\bibinfo{volume}{1305}}, \bibinfo{pages}{055}
  (\bibinfo{year}{2013}), \eprint{1301.7605}.

\bibitem[{\citenamefont{Aprile et~al.}(2012)}]{Aprile:2012nq}
\bibinfo{author}{\bibfnamefont{E.}~\bibnamefont{Aprile}} \bibnamefont{et~al.}
  (\bibinfo{collaboration}{XENON100 Collaboration}),
  \bibinfo{journal}{Phys.Rev.Lett.} \textbf{\bibinfo{volume}{109}},
  \bibinfo{pages}{181301} (\bibinfo{year}{2012}), \eprint{1207.5988}.

\bibitem[{LUX(2013)}]{LUX2015}
\bibinfo{journal}{First Science Results from the LUX Dark Matter Experiment,
  Talk at SURF Gaitskell/McKinsey}  (\bibinfo{year}{2013}).

\bibitem[{XEN(2013)}]{XENON10T}
\bibinfo{journal}{SNOMASS on the Mississippi, SLAC Workshop Talk, XENON}
  (\bibinfo{year}{2013}).

\bibitem[{\citenamefont{Desai et~al.}(2004)}]{Desai:2004pq}
\bibinfo{author}{\bibfnamefont{S.}~\bibnamefont{Desai}} \bibnamefont{et~al.}
  (\bibinfo{collaboration}{Super-Kamiokande Collaboration}),
  \bibinfo{journal}{Phys.Rev.} \textbf{\bibinfo{volume}{D70}},
  \bibinfo{pages}{083523} (\bibinfo{year}{2004}), \eprint{hep-ex/0404025}.

\bibitem[{\citenamefont{Abbasi et~al.}(2009)}]{Abbasi:2009uz}
\bibinfo{author}{\bibfnamefont{R.}~\bibnamefont{Abbasi}} \bibnamefont{et~al.}
  (\bibinfo{collaboration}{ICECUBE Collaboration}),
  \bibinfo{journal}{Phys.Rev.Lett.} \textbf{\bibinfo{volume}{102}},
  \bibinfo{pages}{201302} (\bibinfo{year}{2009}), \eprint{0902.2460}.

\bibitem[{\citenamefont{Aprile et~al.}(2013)}]{Aprile:2013doa}
\bibinfo{author}{\bibfnamefont{E.}~\bibnamefont{Aprile}} \bibnamefont{et~al.}
  (\bibinfo{collaboration}{XENON100 Collaboration}),
  \bibinfo{journal}{Phys.Rev.Lett.} \textbf{\bibinfo{volume}{111}},
  \bibinfo{pages}{021301} (\bibinfo{year}{2013}), \eprint{1301.6620}.

\end{thebibliography}
\end{acknowledgments}

\end{document}